\documentclass[%
 aip,
rsi,
 amsmath,amssymb,
 reprint,%
]{revtex4-1}

\usepackage[caption=false]{subfig}
\usepackage{graphicx}
\usepackage{dcolumn}
\usepackage{bm}
\usepackage{textcomp, gensymb}

\usepackage[utf8]{inputenc}
\usepackage[T1]{fontenc}
\usepackage{mathptmx}
\usepackage{etoolbox}
\usepackage{xcolor}

\makeatletter
\def\@email#1#2{%
 \endgroup
 \patchcmd{\titleblock@produce}
 {\frontmatter@RRAPformat}.
 {\frontmatter@RRAPformat{\produce@RRAP{*#1\href{mailto:#2}{#2}}}\frontmatter@RRAPformat}
 {}{}
}%
\makeatother
\begin{document}

\preprint{AIP/123-QED}

\title[New applications for the world’s smallest high precision capacitance dilatometer and its new stress implementing counterpart.]{New applications for the world’s smallest high-precision capacitance dilatometer and its stress-implementing counterpart.}

\author{R. Küchler}
\affiliation{Max Planck Institute for Chemical Physics of Solids, Nöthnitzer Strasse 40, 01187 Dresden, Germany}
\affiliation{Innovative Measurement Technology Kuechler, Frankenstraße 13, 01309 Dresden, Germany}
 
\author{R. Wawrzy\'{n}czak}
\affiliation{Max Planck Institute for Chemical Physics of Solids, Nöthnitzer Strasse 40, 01187 Dresden, Germany}
 
\author{H. Dawczak-Dębicki}
\affiliation{Max Planck Institute for Chemical Physics of Solids, Nöthnitzer Strasse 40, 01187 Dresden, Germany}
 
\author{J. Gooth}
\affiliation{Max Planck Institute for Chemical Physics of Solids, Nöthnitzer Strasse 40, 01187 Dresden, Germany}
\affiliation{Physikalisches Institut, Universität Bonn, Nussallee 12, D-53115 Bonn, Germany}

\author{S. Galeski}
\affiliation{Max Planck Institute for Chemical Physics of Solids, Nöthnitzer Strasse 40, 01187 Dresden, Germany}
\affiliation{Physikalisches Institut, Universität Bonn, Nussallee 12, D-53115 Bonn, Germany}

\date{\today}

\begin{abstract}
We introduce a new stress dilatometer with exactly the same size and weight as the world’s smallest miniature capacitance dilatometer (height $\times$ width $\times$ depth $=15$~mm$\times{}14$~mm$\times{}15$~mm, weight: $12$~g). To develop this new device, only a single part of the most recently developed mini-dilatometer, the so-called "body", needs to be replaced. Therefore, the new mini dilatometer with an interchangeable body can be used for high-resolution measurements of thermal expansion and magnetostriction with and without large stress. We also report two novel applications of both mini-dilatometer cell types. Our new setup was installed for the first time in a cryogen-free system (PPMS DynaCool). The first new setup allows the rotation of both dilatometers \textit{in situ} at any angle between $-90\degree{}\geq{}\mu\geq{}+90\degree$ in the temperature range from $320$~K to $1.8$~K. We also installed our mini-cells in a dilution refrigerator insert of a PPMS DynaCool, in which dilatometric measurements are now possible in the temperature range from $4$~K to $0.06$~K. Because of the limited sample space, such measurements could not be performed so far. For both new applications, we can resolve the impressive length changes to $0.01$~\AA. 
\end{abstract}

\maketitle

\section{\label{chap1}Introduction}


In many of today's most interesting materials, strong interactions exist between the magnetic moments, electrons, and the underlying crystal structure. These materials can exhibit exciting physical phenomena. Examples include superconductors, magnets, topological insulators, non-Fermi liquids, different forms of quantum criticality, and magnetic frustration. However, because the temperature and energy scales in these materials are typically extremely small, the study of these states and their emergent excitations is highly demanding~\cite{1,2}). 

Ultrahigh-resolution capacity dilatometry is a particularly
suitable technology that is especially sensitive to phase
transitions, very low-energy excitations and the coupling of
multiple electronic, orbital, and spin degrees of freedom to
the lattice.

In the study of single crystals, dilatometry has a decisive advantage over the measurement of the specific heat $C$: depending on the crystal symmetry, different values and even temperature dependencies sometimes result in linear thermal expansion along the various crystal axes. This allows for a more comprehensive study of the crystal structure as complementary directional information can be obtained.
For second-order phase transitions, the initial pressure dependence of the phase transition temperature can be calculated from the ratio of the discontinuities in the volume thermal expansion ($\Delta \beta$) to the specific heat ($\Delta C$) using the Ehrenfest relation:
\begin{equation}
 (dT_C/dp)_{p \rightarrow 0}=V_{mol}T_C \Delta \beta/\Delta C\label{ehrenfest}
 \end{equation}

A similar relation holds for the linear thermal expansion and allows to estimate the uniaxial pressure dependences~\cite{5,6,7}.

Different pressure dependencies may also exist for different phase transitions and crystal axes.

A prominent example is the measurement of the recently discovered multiphase unconventional superconductor CeRh$_{2}$As$_{2}$~\cite{3,4}, where two phase transitions occur within a very small temperature window ($T_{C}=0.26$~K, $T_{0}=0.4$~K). The linear thermal expansion coefficient exhibited a different sign for the two-phase transitions and thus, based on the Ehrenfest relation, an opposite pressure dependence~\cite{4}. The strong positive pressure dependence of $T_{0}$ is in contrast to the strong negative pressure coefficient observed for magnetic order in Ce-based Kondo lattices. This observation is a key to identifying the non-magnetic nature of the second phase at $T_{0}$. 

The advantage of dilatometry, which can be measured along different crystal axes, was most evident in the study of the geometrically frustrated material CeRhSn, where different temperature dependencies could be observed for two complementary crystal axes~\cite{8}. Therefore, we return to it in more detail in the Introduction.

Dilatometric measurements are also "smoking gun" experiments in the study of quantum criticality because the detection of a diverging Grüneisen ratio (ratio of volume thermal expansion and specific heat) proves the existence of a pressure-induced quantum critical point~(QCP)~\cite{9,10,11,12,13}. Moreover, scaling analysis showed that the critical exponent describing the divergent behavior allows to define the nature of the QCP\cite{zhu}. 

Although capacitive dilatometry was developed and used decades ago~\cite{14,15,16,17}, we have only recently made important improvements to the design of dilatometers, which now allow samples of innovative materials to be studied in new environments with unprecedented accuracy~\cite{18,19,20}. These new applications include very large magnetic fields and very low temperatures~\cite{18,20}. In the Editor’s Pick in 2017~\cite{20}, we introduced “The world’s smallest capacitive dilatometer, for high-resolution thermal expansion and magnetostriction in high magnetic fields’’. Despite extreme miniaturization, the capacitive dilatometer can resolve the length changes to $0.01$~\AA. There are other very smart miniaturized dilatometers~\cite{21,22,23,24}, but such an ultrahigh resolution is unprecedented for a capacitive dilatometer of such a small size. 

Because samples of innovative materials are often only a few millimeters or even smaller in size, the absolute length change of the sample at low temperatures tends to be extremely small, making very high resolution of the dilatometer extremely important. Among the various methods for measuring the length change of a sample, capacitive dilatometry stands out due to the remarkably high resolution that can be achieved, of $\Delta{}L=10^{-10}$~m, which exceeds the resolution of other techniques, including x-ray diffraction~\cite{barron}, optical interferometry~\cite{hamann}, or the use of strain gauges~\cite{kabeya} and Piezo-cantilever technology~\cite{24}, by at least one order of magnitude.

Numerous cryogenic devices have limited space for experimental setups. Owing to the small size of our new dilatometer (height $\times$ width $\times$ depth $=15$~mm$\times{}14$~mm$\times{}15$~mm), new applications have been realized for devices with small sample spaces ~\cite{20}. As an important example, our tiny device can be rotated manually in the probe of a commercial Physical Property Measurement System~(PPMS). The super-compact design also enables high-resolution thermal expansion and magnetostriction measurements in a $15.2$~mm diameter tube of a $37.5$T Bitter magnet at the High Field Magnet Laboratory in Nijmegen to a temperature down to $300$~mK~\cite{20}. 

In this study, we report the development of a stress dilatometer that is exactly as small as the previously introduced mini-
dilatometer. For this purpose, only a single part of the most recently developed mini-dilatometer, the so-called "body", needs to be replaced. Therefore, the new mini dilatometer with an interchangeable body can be used for high-resolution measurements of thermal expansion and magnetostriction with and without large uniaxial stress.

Smart devices with piezoelectric actuators that allow \textit{in situ} strain tuning have recently been utilized for electrical resistivity and magnetic ac-susceptibility measurements on unconventional superconductors~\cite{25,26,27}. However, thermal expansion and magnetostriction are the most sensitive thermodynamic probes of phase transitions, depending on the uniaxial pressure derivatives of entropy and magnetization, respectively, and are even better suited for investigating \textit{ for example } frustrated Kondo lattices~\cite{28}.

A few years ago, we developed the first uniaxial stress-capacitance dilatometer for high-resolution thermal expansion and magnetostriction~\cite{19}. Back then, our dilatometers were already miniaturized but not as small as they are today. Our first stress dilatometer had a much larger cell size (height $\times$ width $\times$ depth $=36$~mm$\times26$~mm$\times20$~mm)~\cite{19}. The operating mechanism and basic design were the same as those of our new mini-stress dilatometer. Using this larger stress cell, a force of up to $75$~N could be applied to the sample. This corresponds to a uniaxial stress of $3$~kbar obtained by measuring cuboid samples with a $0.25$~mm$^{2}$ cross-section. 

In an experiment on CeRhSn single crystals, we successfully utilized uniaxial stress as a control parameter for tuning a quantum critical geometrically frustrated material into a magnetically ordered ground state~\cite{29}. In contrast to all previously studied quantum critical materials~\cite{9,10,11,12,13}, the expansion coefficient $\alpha{}/T$ diverges only within the $ab$ plane, while along the $c$-axis, it displays Fermi-liquid behavior~\cite{8}. This has been explained by a QCP related to geometrical frustration, which is sensitive only to in-plane stress, but not to $c$-axis stress~\cite{8}. 
 
We then utilized a newly developed uniaxial stress dilatometer~\cite{19} for further investigation~\cite{29}. We applied various uniaxial stresses up to $2$~kbar on a well-characterized single rectangular crystal with the same cross-section at both ends, which guarantees uniform stress along the sample. The applied stress was enhanced by reducing the cross section of the samples. The sample with the highest stress of $2$~kbar had a cross-section of $0.5\times{}0.6$~mm$^{2}$.

\begin{figure}
\subfloat{\label{fig1a}}
\subfloat{\label{fig1b}}
 \centering
 \includegraphics[width=\linewidth]{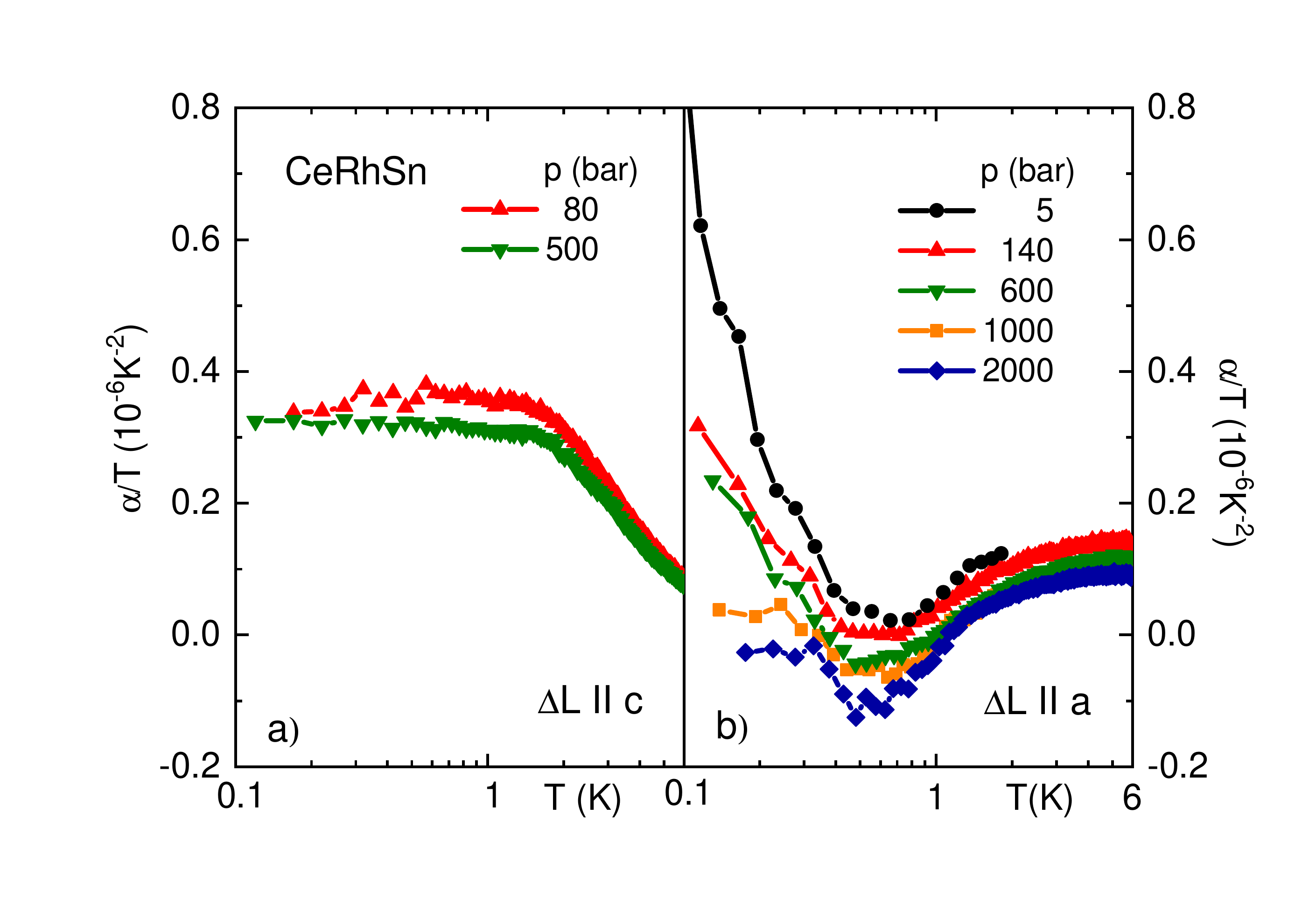}
 \caption{Thermal expansion coefficient divided by temperature $\alpha/T$ measured along the $c$- and $a$-axis as a function of temperature at different uniaxial stress applied parallel to the measurement direction. The data at $0.5$~MPa were taken from Ref.~\onlinecite{8}.}
 \label{fig1}
\end{figure}

As can be seen in Fig.~\ref{fig1b}, increasing the stress within the Kagome plane leads to a suppression of the low-temperature divergence in $\alpha{}/T$, followed by a step-like change in the expansion coefficient, indicating a second-order phase transition. 

Because Kondo coupling increases with stress, which alone would stabilize the paramagnetic behavior in CeRhSn, the observed order arises from the release of geometrical frustration by the in-plane stress. Thus, uniaxial stress has been successfully utilized as a novel control parameter for tuning a quantum-critical geometrically frustrated material into a magnetically ordered ground state.

Our study on CeRhSn with strain-dilatometry measurements is one of the first of this type and opens a different way to investigate geometrically frustrated matter. It is expected that, in several other quantum spin liquid candidate materials, different hidden quantum phases can be discovered by the release of frustration with uniaxial stress.

Using our newly developed tiny stress cell, a similar force of up to $65$~N can be applied to the sample. This corresponds to a uniaxial stress of $2.6$~kbar obtained by measuring cuboid samples with a $0.25$~mm$^{2}$ cross-section. The mode of operation and design of the new mini-stress dilatometer are explained in detail in Sections ~\ref{chap2a}~and~\ref{chap2b}. 

In this study, we report two novel applications, in which both mini-dilatometer cells can be used. Our novel ultra-high-resolution dilatometry setups were installed for the first time in a cryogen-free system (PPMS DynaCool). This dry system uses a single, two-stage pulse-tube cryocooler for both the superconducting magnet and temperature control system, providing an efficient, low-vibration environment for sample measurements. 
 
Our first new setup was a PPMS dilatometry probe that included an \textit{in situ}  mechanical rotator. Therefore, it is now possible to rotate both mini-dilatometers  \textit{in situ} at any angle between $-90\degree{}\geq{}\mu\geq{}+90\degree$. \textit{In situ} means that the dilatometer remains inside the insert during rotation and does not have to be removed beforehand. This rotation can be performed at temperatures between $1.8$~K and $320$~K. This is reported in Section ~\ref{chap3}. 

We also installed our mini-cells in a dilution refrigerator insert of a PPMS DynaCool, which enables access to a temperature range spanning $4$~K all the way down to $60$~mK. The space for user experiments in such a dilution refrigerator is only $22$~mm in diameter by $35$~mm long cylindrical volume. Here, we mounted the cells parallel and perpendicular to the applied magnetic field. This is the first time dilatometry measurements have been performed in a PPMS dilution refrigerator. The cooling of the dilatometer to $60$~mK can be achieved very quickly in less than $8$ h. Detailed information on the experimental setup and first measurement results are presented in Chapter ~\ref{chap4}. 

For both new applications, we can resolve the impressive length changes down to $0.01$~\AA, which is $10$ times smaller than that reported previously for dilatometry measurements in PPMS systems~\cite{20}. This corresponds to a relative resolution of $\Delta{}L/L=10^{-10}$ obtained by measuring a sample with a length slightly smaller than 1 mm. 

Until recently, ultra-high-resolution measurements of $\Delta{}L/ L=10^{-10}$ could only be performed using a dilatometer inside the inner vacuum chamber in the dilution refrigerator of an Oxford Instruments cryostat~\cite{18,19,20}. By using Oxford Instruments cryostats, we made great efforts to mechanically disconnect the cryostat from its environment using sophisticated self-built damping systems.

Our new results prove that commercial cryogen-free systems (PPMS DynaCool) are also well suited for ultrahigh-resolution thermal expansion and magnetostriction measurements using our cells, which are relatively insensitive to mechanical vibrations.

The number of cryogen-free systems in laboratories worldwide is steadily increasing as liquid helium becomes a less available resource. These types of systems are expected to gradually replace older cryogenic systems. Because dilatometer measurements are extremely sensitive to mechanical or electrical disturbances, the ultra-high-resolution measurements we have achieved promise that other sensitive applications can be operated in cryogen-free systems as well. 

However, to achieve such high-resolution measurements, it is necessary to set up the PPMS system in such a way that mechanical and, above all, electronic noise are avoided. Here, avoidance of ground loops is essential.

This is the first time we have achieved exceptionally good resolution using a PPMS system. When setting up the DynaCool measuring station, special emphasis was placed on avoiding ground loops. This prompted us to review the setup of the cryogenic PPMS system. After we were able to eliminate the existence of ground loops in conventional cryogenic PPMS-systems, we were also able to significantly improve the resolution.

Because the avoidance of electronic noise is of particular importance for the success of dilatometric measurements, we will discuss our DynaCool experimental setup in detail in Chapter ~\ref{chap3a} and report on how we have improved our other PPMS systems to reduce electronic noise.
In Chapter ~\ref{chap5}, we present the first test measurements using the new mini-stress dilatometer.

\section{\label{chap2}The new Mini-Stress-Dilatometer}
 \subsection{\label{chap2a}The new Mini-Stress-Dilatometer}

\begin{figure}
 \centering
 \includegraphics[width=\linewidth]{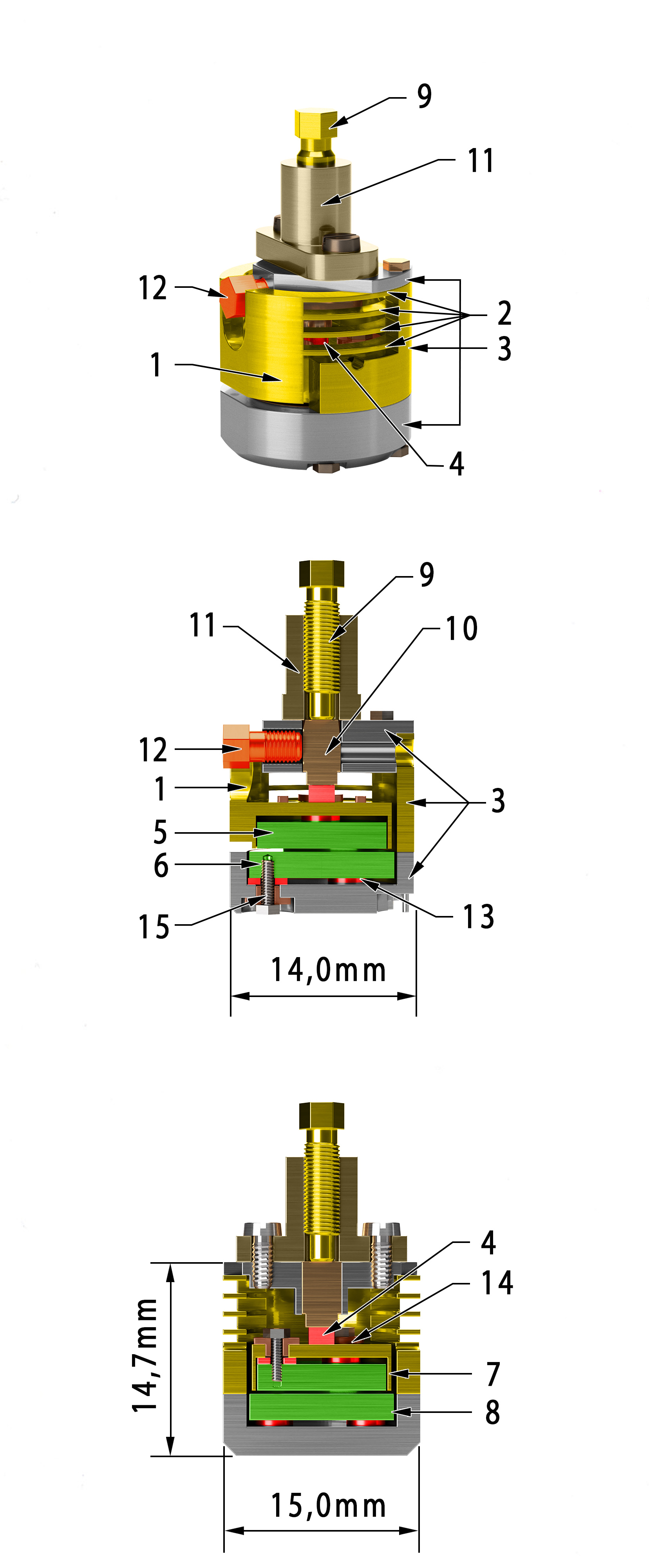}
 \caption{Schematic drawing of our new mini-stress-dilatometer. The top picture shows a 3D-view, the middle picture shows a side cut-away view, and the one at the bottom shows a front cut-away view of the dilatometer. (1) Moving part, (2) four Be–Cu flat leaf springs, (3) external frame, (4) sample, (5) upper capacitor plate, (6) lower capacitor plate, (7 and 8) guard rings, (9) adjusting screw, (10) cubic piston, (11) removable sample- adjusting-tool, (12) locking screw, (13) sapphire-washer, (14) insulating piece of vespel, and (15) electrical connection soldered on the screw.}
 \label{fig2}
\end{figure}

\begin{figure}
 \centering
 \includegraphics[width=0.9\linewidth]{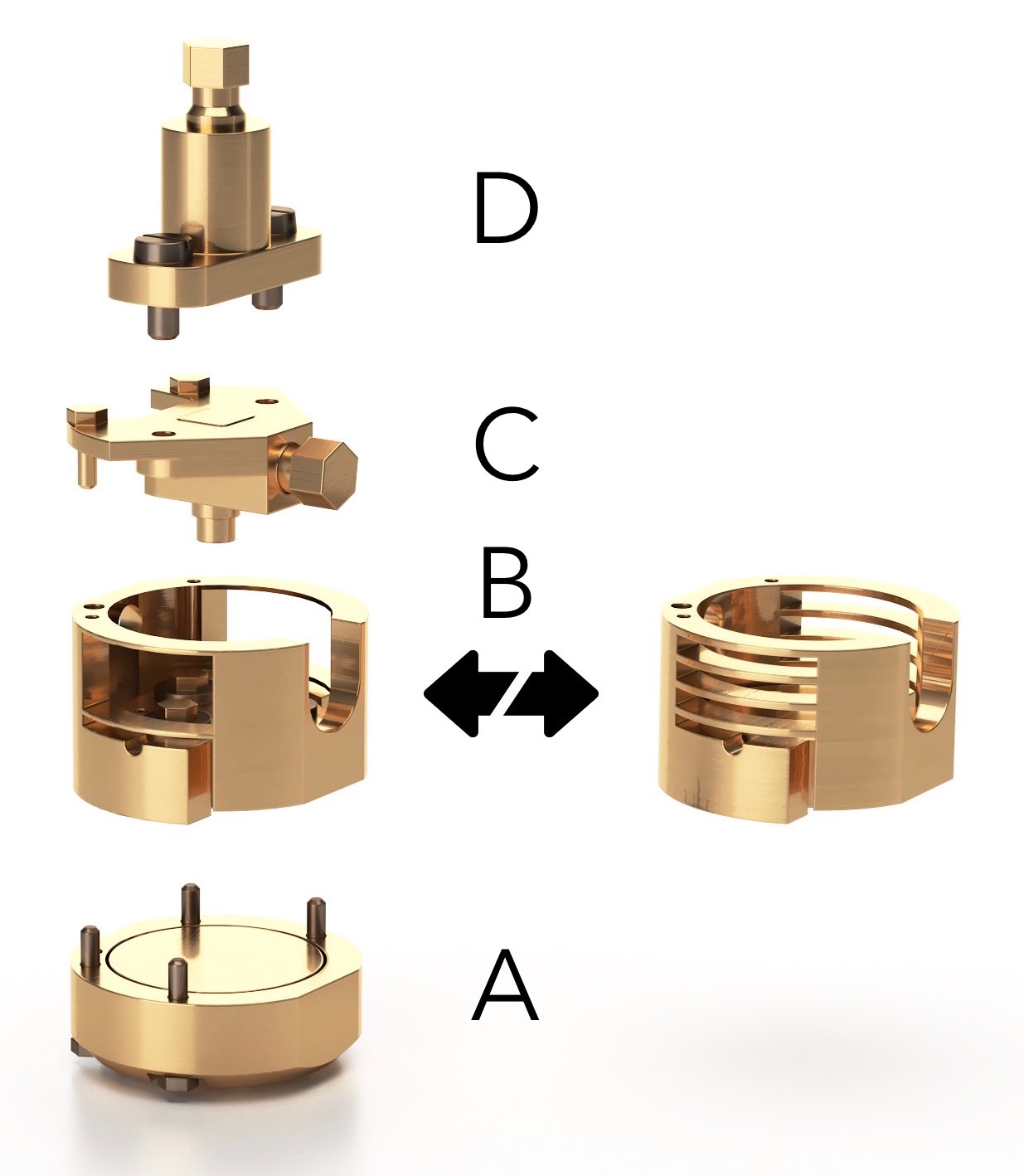}
 \caption{A comparison of the new mini-stress dilatometer with the almost zero pressure mini-dilatometer: Each dilatometer consists of four main parts: (A) Bottom part, (B) Main body, (C) Cover and (D) Sample-adjusting tool. To apply additional uniaxial stress, just the main body (B) has to be exchanged. The stress cells in the main body contain four springs with a thickness of $0.5$~mm instead of two springs with a thickness of $0.25$~mm.}
 \label{fig3}
\end{figure}

The mini-stress-dilatometer is based on our patented mini-dilatometer design~\cite{18,19,20}, which is in line with the Pott-Schefzyk principle~\cite{16} of two flat parallel leaf springs. In contrast to this well-known principle, our new stress cell uses four springs instead of two. The new design is shown schematically in Fig. ~\ref{fig2}. Our stress cell consists of an external frame (3) and a moving part (1). The lower capacitor plate (6) is mounted on the lower part of the external frame (3). The upper capacitor plate (5) is fixed at the bottom of the moving part (1). The external frame and moving part are connected by four flat leaf springs (2) with a thickness of $0.5$~mm. The sample is clamped using an adjusting screw (9), which presses the sample against the force of the four parallel leaf springs. In this construction, a change in the length of the sample causes an equally large change in the length between the two capacitor plates. While the Pott-Schefzyk dilatometer consists of many individual parts, our design consists of only four main parts: (A) bottom part, (B) main body, (C) cover, and (D) sample-adjusting tool (see Fig. ~\ref{fig3}). 

The sample-adjusting tool (D) was exclusively used for sample mounting and was not used during measurement. The bottom part (A) contains the lower capacitor plate (6). The main body (B) contains flat leaf springs and an upper capacitor plate (5). This main body (B) is responsible for cell function. In the Pott-Schefzyk cell, the main body was made up of $7$ different parts, which were screwed together. In our mini-dilatometer, the main body was manufactured as a single part by wire erosion and milling. If the main body of the new stress dilatometer was manufactured using the Pott-Schefzyk design, $13$ individual parts would have been required.  Compared with the original mini-dilatometer cell~\cite{20}, two fundamental changes were made to this main body (B): (i) The thickness of the leaf springs increased significantly from $0.25$~mm to $0.5$~mm. Our theoretical calculations showed that the spring force $F$, and accordingly, the applied uniaxial pressure (stress) to the sample $p$, increases with increasing spring thickness $d$ as $F\propto{}p\propto{}d^{3}$. Consequently, the spring force on the sample will increase $8$ times by doubling the spring thickness. (ii) Two additional springs were included in the parallel spring circuit. In the original circuit, the two springs are parallel because they are connected side-by-side to act as a single spring. The strain of the ensemble is the common strain, and the stress of the ensemble is the sum of their stresses. The same holds true when more than two springs are parallel. We included two additional leaf springs with the same $0.5$~mm thickness to our dilatometer design. Consequently, the ensemble stress should be doubled. Combining both fundamental changes ((i) and (ii)), the substantially stronger spring force acting on the sample should be $16$ times larger than that of the mini-dilatometer design. In the following section, we describe the experimental determination of the resulting spring force acting on the sample. This shows that our calculations were considerably precise. In fact, the spring force increases by $15$ instead of $16$. 

The third main part of the dilatometer is the cover (C), which is screwed onto the body. This cover includes a lock screw (12) whose function is described below. Part number four is the sample-adjusting tool (D), which is used to mount the sample and will be removed afterwards. This sample-adjusting tool included a fine-threaded adjusting screw (9), which was used to mount the sample (4, red). The sample is easily mounted as follows: First the sample-adjusting tool (D) and cover (C) are unscrewed and removed. Next, the sample is inserted into the center of the body from above. The cover (C) and adjustment tool (D) are then screwed back on. The sample is initially free-standing and is clamped in the next step by tightening the adjusting screw (9). In this method, the screw does not press directly on the sample but is placed on a cubic piston (10) that can only move horizontally in the lid. This ensures that the intended orientation of the single crystals does not change during clamping. Once the sample is clamped, a locking screw (12) is used to fix the cubic piston (10). The sample-adjusting tool (D) is then unscrewed and removed. 

The mini-dilatometer and the mini-stress dilatometer are
almost identical in construction. Three of the four main parts (bottom part (A), cover (C), and sample-adjusting tool (D)) can be used for both dilatometers (see Fig. ~\ref{fig3}). Only the main body (B), which contains the leaf springs, has to be exchanged for applying “almost- zero’’ or significant uniaxial pressure to the sample. Therefore, the mini-stress dilatometer has exactly the same tiny size and weight as the world’s smallest miniature dilatometer (height $\times$ width $\times$ depth $=15$~mm$\times{}14$~mm$\times{}15$~mm, weight: $12$~g). Both dilatometers can be used to measure samples smaller than $1$~mm to $2.75$~mm in size. In the case of a mini-dilatometer, samples with any geometry can be installed. In contrast, for the stress dilatometer, well-characterized rectangular single crystals with the same cross section at both ends are required to ensure a uniform pressure distribution within the sample. Compared to recently developed piezo strain tuning devices~\cite{27}, our stress dilatometer has the advantage that, for rectangular crystals, there is no strain gradient within the sample. When applying a higher pressure by measuring samples with smaller cross-sections, it is advisable to polish the sample surfaces well to prevent cracks.

All dilatometer parts (except for the electrically insulating washers) were produced from ultrapure Be–Cu to reduce the eddy currents induced by the time variation of the magnetic field. A Be component of $1.84$\% produces an electrical conductivity that is much lower than that of pure copper or silver.

A commercial capacitance measuring bridge can be used to record the measured change in capacitance; the formula of the plate capacitor can then be applied to calculate this change in capacitance back to the length change caused by the sample. The distance between the capacitor plates of the unloaded cell is $0.25$~mm ($C=3$~pF), whereas at the measuring position, it decreases to approximately $0.05$~mm ($C=20$~pF). All the mini-stress dilatometers manufactured cause a short circuit at a capacitance exceeding $40-50$~pF. This makes it possible to work with a very high measuring capacitance between $10$ and $20$~pF: the absolute value of the capacitance is measured by a commercial capacitance measuring bridge (Andeen Hagerling 2550A) with a resolution of $10^{-6}$~pF. At such a high capacitance, the absolute length change of the sample can be measured with a sensitivity of $\Delta{}L=0.01$~\AA. Despite their small dimensions, the new mini-stress dilatometers are extremely high-resolution dilatometers. Our new design also allows measurements under substantial uniaxial stress. In the next chapter, we will explain a force-testing facility that we set up to determine the spring force exerted on the sample.
 
 \subsection{\label{chap2b}Spring force and resulting stress exerted on the sample.}

The new mode of operation of the mini-stress dilatometer is based on the enormous force exerted on the sample by four parallel leaf springs. We mentioned in Ref.~\onlinecite{20} that in our mini-dilatometer design with two $0.25$~ mm-thick leaf springs, a small spring force between $3$ and $4$~N was applied to the sample. In most experiments, such a weak force did not cause any changes in the material properties. In contrast, in our new design with $4$ thicker and parallel springs, the spring force, and consequently the induced uniaxial pressure (stress), increases significantly. In the following, the applied spring force is experimentally determined for the mini-dilatometer with two $0.25$-mm-thick springs and two mini-stress dilatometers with four $0.4$-mm-thick and $0.5$-mm-thick springs, respectively.

The following formula allows the calculation of the change in distance between the plates of the corrected plate capacitor:

\begin{equation}\label{eq1}
 \Delta{}L=\epsilon_{0}\pi{}r^{2}\frac{C-C_{0}}{CC_{0}}\left(\frac{1-CC_{0}}{C_{\textrm{max}}^{2}}\right).
\end{equation}

We used Equation~\ref{eq1} to test the functionality of the different mini-dilatometers at room temperature and to determine the spring force exerted on the sample. Here, $\Delta{}L$ is the length change of the sample, \textit{i.e.}, the distance change between the capacitor plates. $C$ is the changing capacitance and $C_{0}$ is the initial capacitance. In contrast to the ideal plate capacitor, the capacitor plates of the manufactured dilatometer do not have an exact plane-parallel orientation. This error owing to tilting is considered in the formula used by including the short-circuit capacitance $C_{\textrm{max}}$~\cite{16,20}. We obtained the short-circuit capacitance by carefully decreasing the plate distance with an adjustment screw until the capacitor was shortened. The final and highest measured values were $C_{\textrm{max}}$. Considering the tilting of the plates, Pott and Schefzyk~\cite{16} derived Equation~\ref{eq1}, which is a corrected expression for the measured length change $\Delta{}L$, where $\epsilon_{0}=8.8542\times10^{-12}$~F/m is the permittivity in vacuum and $r=5$~mm is the radius of the circular smaller upper capacitor plate. 

\begin{figure}
\subfloat{\label{fig4a}}
\subfloat{\label{fig4b}}
 \centering
 \includegraphics[width=0.4035\linewidth]{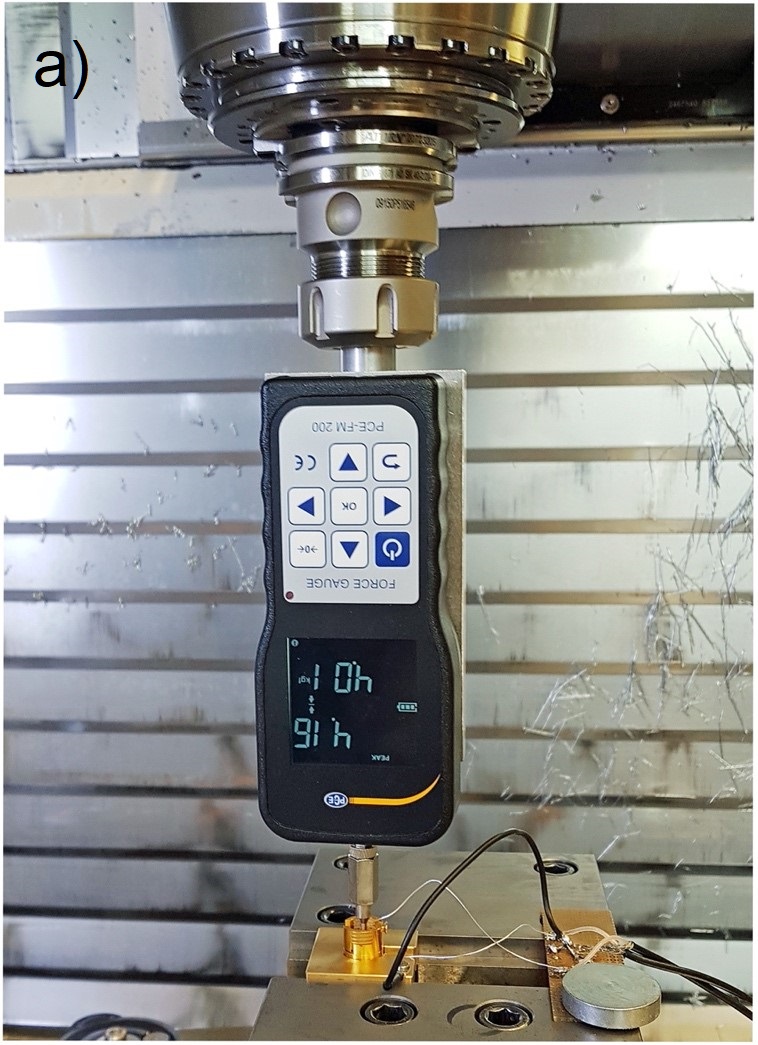}\includegraphics[width=0.5965\linewidth]{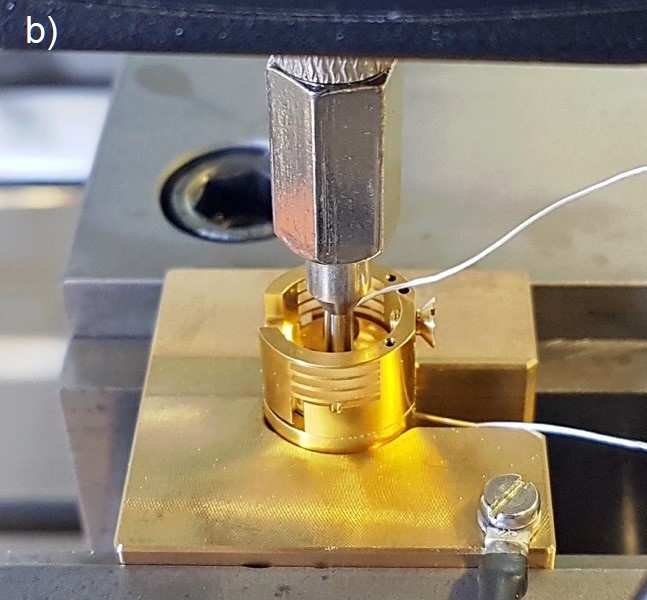}
 \caption{Determination of the resulting spring force: (a) A force gauge is mounted on a CNC machine. (b) The sensor tip of the force gauge is set perpendicular to the upper surface of the moving part of the dilatometers. When the force is increased up to $7000$~g, the tip increases its pressure on the moving part and causes a downward vertical movement of the moving part, including the upper capacitor plate, towards the lower plate.}
 \label{fig4}
\end{figure}

Fig.~\ref{fig4} shows pictures of our experimental setup, which we used to determine the spring force applied to the sample as a function of the capacitance. A force gauge with a resolution of $0.01$N is used to measure the applied force. The force gauge was mounted on a CNC machine that allowed precise vertical movement in micrometer steps. The sensor tip of the force gauge acted centrally on the moving part of the dilatometer (see Fig. ~\ref{fig4}). Using remote control of the CNC machine, we slowly moved the force gauge together with the sensor tip downwards. Thus, we gradually increased the force that led to stronger bending of the springs and reduced the distance between the two capacitor plates. In this way, we determined a calibration curve $F(N)$ vs. $C(pF)$ by increasing the force stepwise by $2$~N for the stress mini-dilatometer and by $0.2$~ N for the mini-dilatometer. The change in capacitance was simultaneously measured using an Andeen Hagerling~(AH) 2550A capacitance bridge. The resulting change in length $\Delta{}L$ (Distance between the capacitor plates) was calculated from the measured capacitance using Equation~\ref{eq1}. 

\begin{figure}
 \centering
 \includegraphics[width=\linewidth]{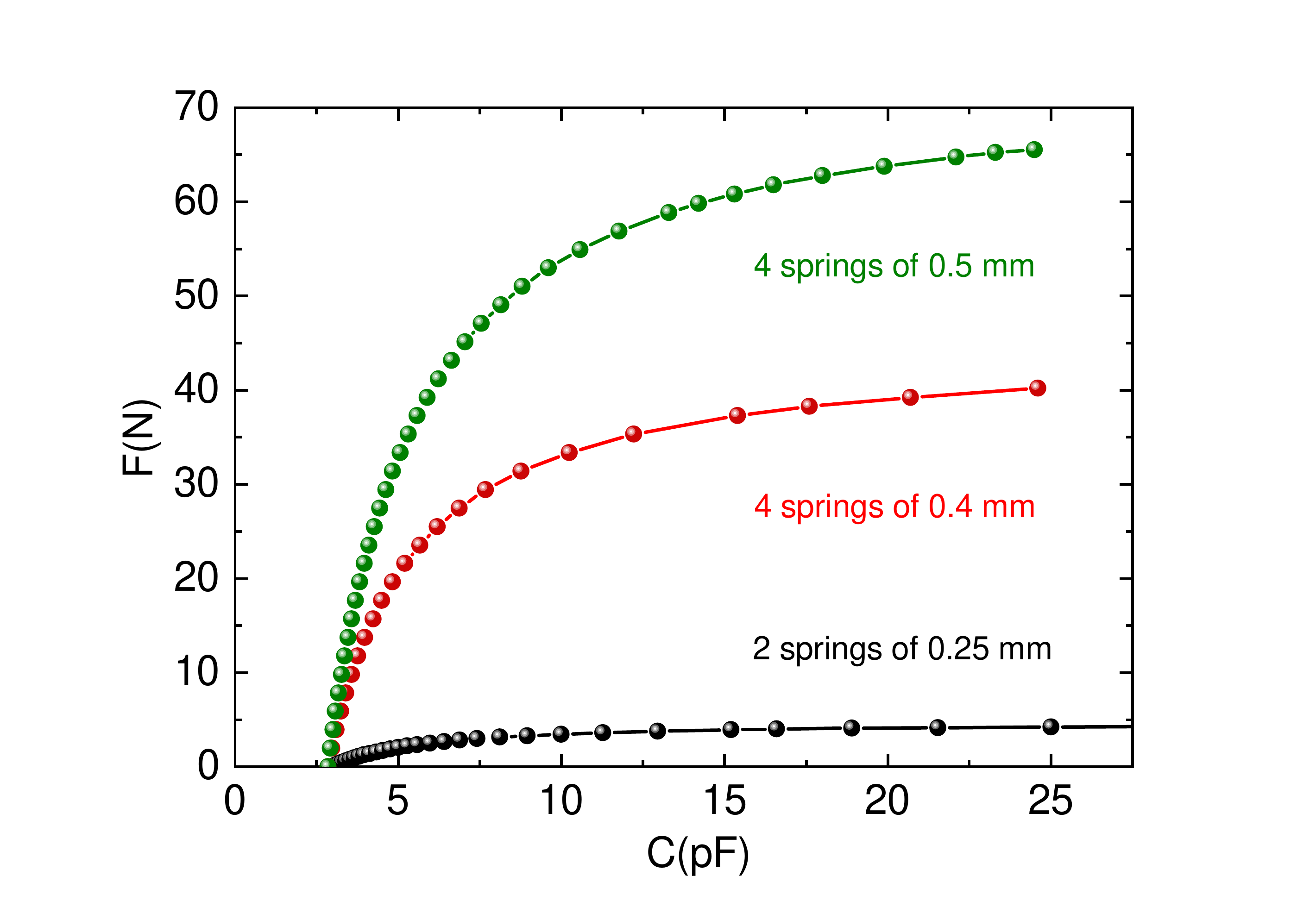}
 \caption{Capacitor plate displacement and respective spring force as function of working capacitance. We typically operate the dilatometer in between $10$ and $20$~pF corresponding to forces between $55$ and $65$~N for the strongest stress-cell indicated by dotted lines.}
 \label{fig5}
\end{figure}

Fig.~\ref{fig5} shows the relationship between the measured capacitance and the spring force acting on the sample for all three dilatometers. As the dilatometers are operated between $10$ and $20$~pF, we obtained the expected force of $3$ to $4$~N for the mini-dilatometer ($2$ springs of $0.25$~mm thickness). Because the resolution of the dilatometer improves in square form with increasing capacitance, high-resolution measurements are not obtained when using a measuring capacitance of less than $10$~pF. For the two stress-dilatometer with $4$ springs of $0.4$~mm and $0.5$~mm thickness, the spring force significantly increased by approximately $9$ and $15$ times, respectively. For a stress cell with $4$ springs of $0.5$~mm thickness, the spring force in the operating range is between $50$ and $65$~N. This corresponds to a maximal uniaxial stress of $0.65$ and $4$~kbar obtained by measuring a cuboid sample with $1$~mm$^{2}$ and $0.4$~mm$^{2}$ cross sections, respectively. As the modulus of elasticity of copper alloys remains almost constant from room temperature to below $1$~K, this value applies to the entire temperature range from $320$~K to $0.01$~K. 

\begin{figure}
 \centering
 \includegraphics[width=\linewidth]{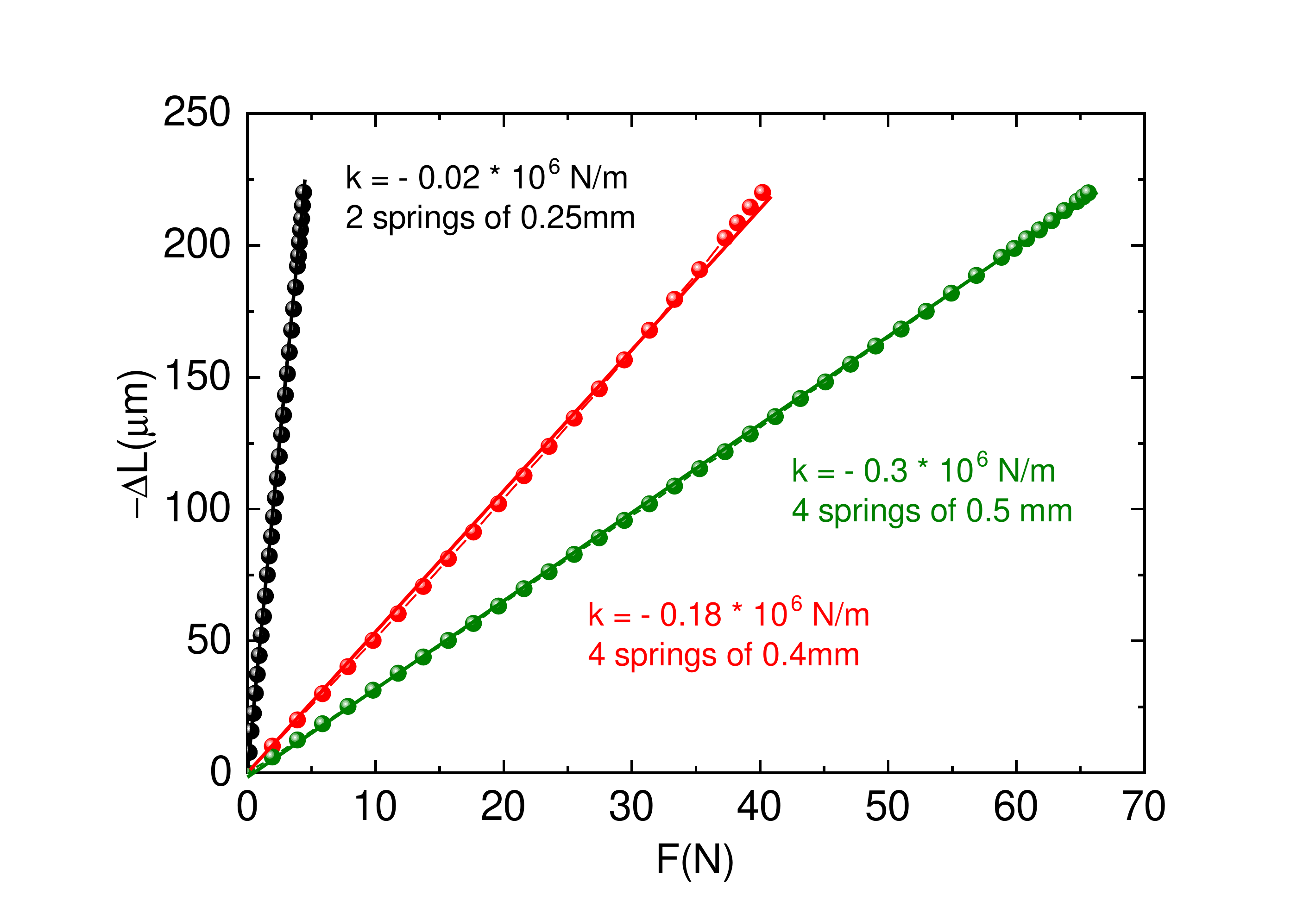}
 \caption{Experimentally determined relation between the applied spring force and the displacement of the upper capacitor plate from its rest position, $250$~$\mu$m above the lower plate. The dotes lines display Hooke’s law $F=k\times{}x$ with spring constant $k$ as quantified.}
 \label{fig6}
\end{figure}

Fig.~\ref{fig6} shows the obtained linear relation between the measured length change $\Delta{}L$ (calculated using Equation~\ref{eq1}) and the applied force. The observation of a linear relationship $F= k \times \Delta{}L$, which is valid here for all three dilatometers, proves that the springs do not plastically deform. The bending of the springs is still in the elastic range even at the highest applied force. This means that also the mini stress dilatometer with $4$ springs of $0.5$~mm thickness and with the largest spring constant of $k=-0.3\times{}10^{6}$~N/m can be used over the entire working range.

\section{\label{chap3}New application: \textit{In situ} dilatometry probe for the PPMS system}

All miniature capacitive dilatometers developed worldwide to date have a size that would ideally fit into a PPMS system\cite{18,19,21,22,23,24}, but exclude the rotation of the dilatometer. The only exceptions are our mini dilatometer and the dilatometer developed by Quantum Design~(QD). In this QD design~\cite{30}, the dilatometer was located inside a capsule. The capsule is configured to allow the dilatometer cell to be rotated and locked in place via a set screw to accomplish measurements at various angles. However, the dilatometer must be removed from the PPMS before each rotation, and the angle must be reset manually. \textit{In situ} rotation at low temperatures is not possible with this method. In addition, in this application, it is preferable that the sample to be measured is $2$~mm in length, $\pm{}50$~$\mu$m. 

In the following, we present a newly developed dilatometer probe that allows the dilatometer to be rotated \textit{in situ} at any angle between $-90\degree{}\leq\mu\leq{}+90$ in PPMS over a temperature range from $320$~K to $1.8$~K. This allows for systematic anisotropy studies with short measurement times. Both the mini- and mini-stress dilatometer can be used in this application to measure samples less than $1$~mm to $2.75$~mm in size.

\begin{figure}
\includegraphics[width=0.95\linewidth]{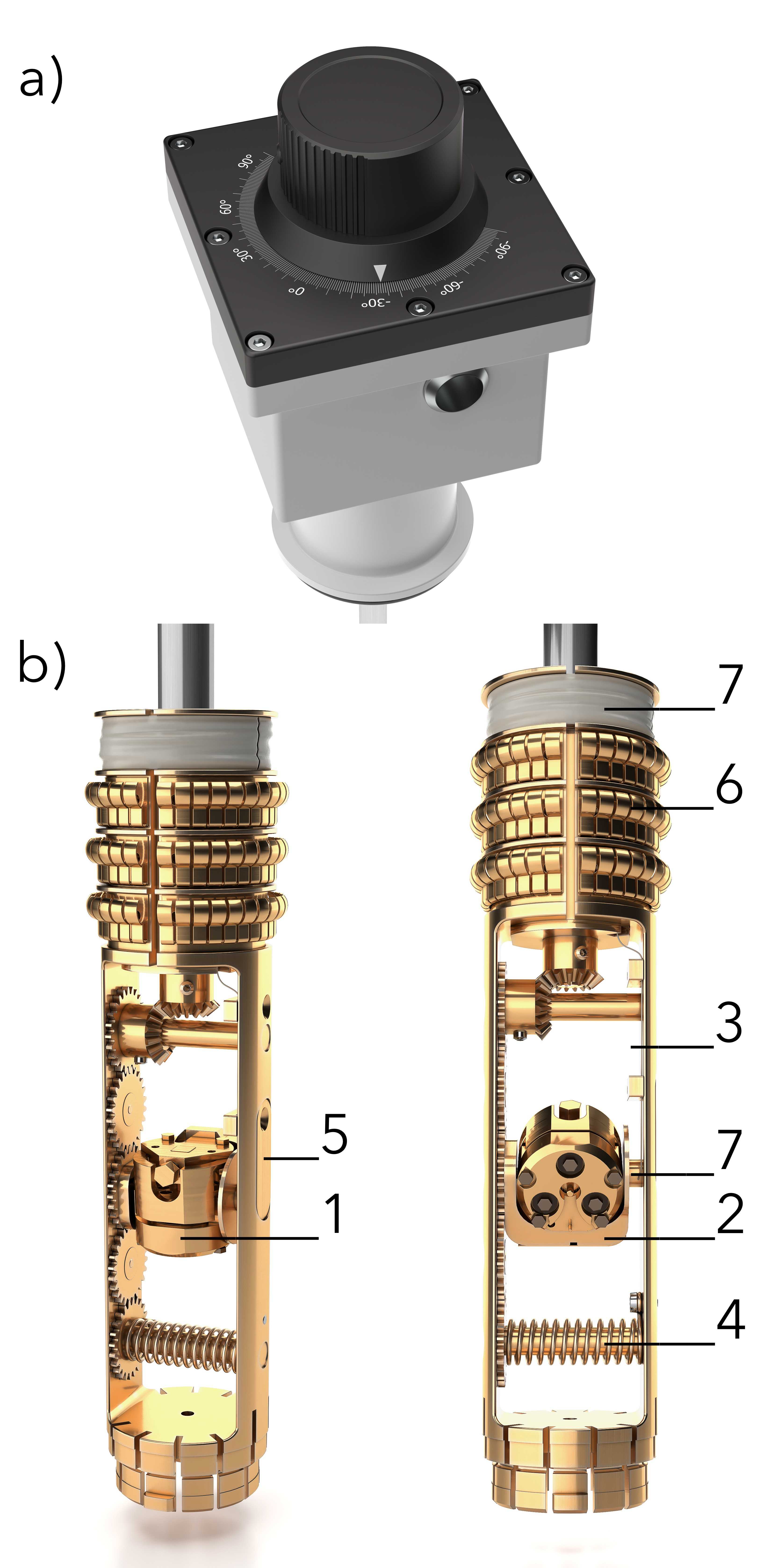}
 \caption{(a) Head of the probe and (b) In-situ PPMS-dilatometry probe from two different views. (1) Dilatometer, (2) C-shaped dilatometer holder, (3) Cage, (4) Spiral spring, (5) Leaf spring, (6) Gold-plated contact springs, (7) Position of the Coax-cables.}
 \label{figc3}
\end{figure}

Fig. ~\ref{figc3} shows the developed PPMS dilatometry probe. The mini-dilatometer (1) or, alternatively, the mini-stress dilatometer is mounted to a C-shaped dilatometer holder (2) placed on a mechanical rotator that allows horizontal manual \textit{in situ} rotation of the sample in the cell within the PPMS. The axis of rotation of the dilatometer is perpendicular to the direction of the applied field.

A probe head made of anodized aluminum (see Fig. ~\ref{figc3}(a)) was attached to the upper end of the probe stick. This head contained a rotary knob with which the dilatometer could be manually rotated between $-90\degree$ and $+90\degree$. We included a locking mechanism to prevent the rotation of more than $180\degree$. This is to prevent the ultrathin coaxial cables used for wiring from overtwisting and breaking. However, for special experiments and care, the pin of the locking mechanism can be removed to allow rotation of up to $360\degree$.

The rotary knob, fixed on the probe head, is screwed to a stainless-steel tube, which goes down to the cage (3) via a hermetic feed-through. A cage (3) is screwed to the lower end of the tube, which contains the mechanical rotator and dilatometer holder, including the dilatometer. All parts of the mechanical rotator are made with very high precision in our workshop from bronze, which allows operation in very strong magnetic fields (current max. tested field: $12$~T). The lower end of the stainless-steel tube is screwed on a bevel gear wheel, which transmits rotation to the C-shaped dilatometer holder (2) via another bevel gear wheel and a set of gear wheels. The lowest gear wheel is attached to a spiral spring (4), which is under tension and thus prevents backward movement. This enables precise positioning of the dilatometer. A leaf spring (5) is mounted in the middle of the outer cage frame, which exerts a slight pressure on the C-shaped dilatometer holder and center gear wheel to optimize the thermal coupling of the dilatometer to the cage.
 
The PPMS-probe is thermally coupled to the annular region at the bottom of the PPMS via contact springs, where the heaters warm the helium gas to the correct temperature via a pin connector. Three rows of gold-plated contact springs (6) are attached to the top of the cage. The thermal anchors (6) are mounted directly above the dilatometer and touch the lower part of the inner chamber of the PPMS cooling channel. Only the lower part of the inner PPMS cooling channel is made of a highly heat-conductive material (copper) and maintains the same temperature as the pin connector. The reason the thermal anchors’ work platform with its extra-large surface is mounted at this level is to effectively improve the thermal coupling of the mini-dilatometer. Because the coaxial cables also provide heat input, they are wrapped several times in the row directly above the contact springs (7) to dissipate this additional heat to the cooling channel. The sample chamber should also be kept under a helium atmosphere with a typical helium pressure of 5-7 Torr, which further improves the thermal stability of the cell and the sample inside. The probe also contains radiation shields to prevent additional heating.

To check the thermal coupling of the dilatometer, we screwed a Cernox-CX-SD thermometer onto the mini-dilatometer and then measured the temperature during warm-up directly on the dilatometer and on the PPMS (pin-connector). When the temperature increased from 2 K to 300 K at a sweep rate of 0.3 K/min, the difference between the two thermometers was less than 0.2 K for temperatures up to 50 K, which increased to 0.5 K up to 100 K, and then remained nearly constant until 300 K. When we started at higher temperatures, the temperature difference at the thermometers remained less than 0.2 K for a large temperature window. When sweeping with 0.1 K/min. we did not observe a temperature difference below 50 K between the thermometers.    

The head of the probe also contained two hermetically sealed LEMO connectors, which were connected to a capacitance bridge with a pair of coaxial cables. 

\begin{figure}
 \centering
 \includegraphics[width=\linewidth]{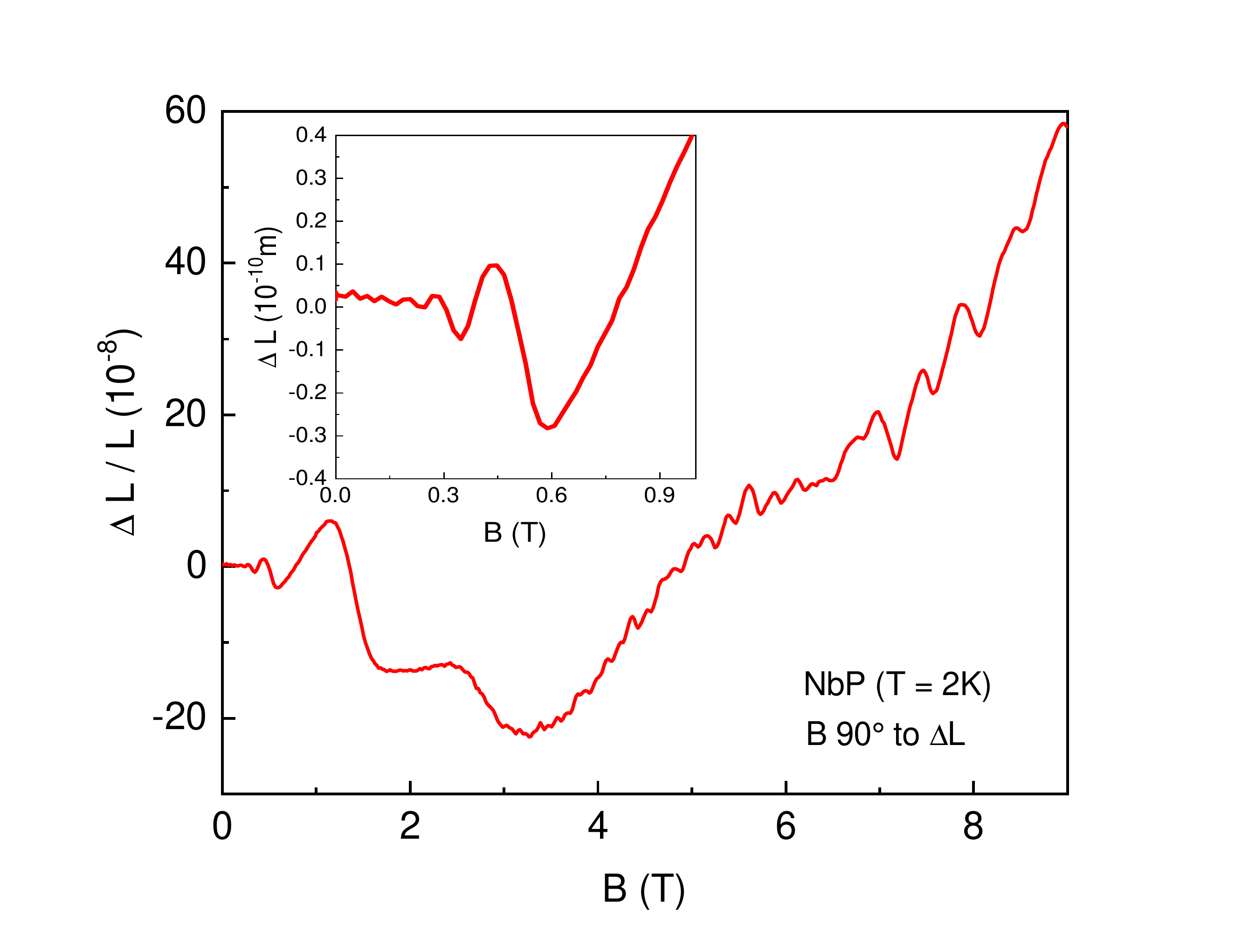}
 \caption{The relative change in length $\Delta{}L/L$ of a $1$~mm large NbP single crystal at $2$~K is shown as a function of the applied magnetic field; quantum oscillations in the length change due to the de Haas-van Alphen effect can clearly be seen. The field was swept in a DynaCool system with $10$~Oe/s.}
 \label{fig7}
\end{figure}

As an example of the exceptional resolution of our \textit{in situ} PPMS dilatometry probe operating in a DynaCool system, we show the quantum oscillation measurements in the magnetostriction $\Delta L(B) /L$ of a Weyl semimetal NbP (see Fig. ~\ref{fig7}). A $1$~mm single crystal was mounted so that the change in the length $\Delta{}L(B)$ of the sample is measured along the $c$-axis of the crystal, and by rotating the dilatometer to $\theta=90\degree$, the magnetic field $B$ is applied along the $a$-axis. The field was swept at $2$~K in the DynaCool system at $10$~Oe/s. Clear de Haas van Alphen~(dHvA) oscillations become visible at very low fields and reach very strong amplitudes at the highest fields measured. The magnetostriction in NbP arises because the carrier concentration is changed by the magnetic field. Magnetostriction measurements are very useful to study semimetals with light carriers and multiband contributions to the density of states~(DOS)~\cite{31}. In contrast to transport~\cite{32}, the thermodynamic probe is directly coupled to the DOS because magnetostriction simply measures the linear function of the field-induced change in carrier density $\Delta{}L/L=c\Delta{}N(B)$~\cite{31,33}. The applied magnetic field quantizes the energies of the quasiparticles into Landau levels, which occupy discrete energy levels perpendicular to the magnetic field direction, depending on the cyclotron energy~\cite{34}. As the cyclotron energy increases with increasing field, higher Landau levels become depopulated. The ripples and peaks observed in Fig. ~\ref{fig7} are caused by sudden changes in carrier concentration when the Landau level is evacuated, and they reflect asymmetric singularities in the DOS.

In NbP, the observed peaks were associated with two-electron and two-hole Fermi surface pockets~\cite{35}. As shown in the inset in Fig. ~\ref{fig7}, low-frequency quantum oscillations can be resolved from as low as $0.3$~T. This corresponds to a large magnetic length, which characterizes the high quality of the sample and the ultrahigh sensitivity of our instrument. Owing to the high sensitivity of our setup, displacements as small as $\Delta{}L=0.01$~\AA can be resolved. Our magnetostriction measurements of NbP, among the first on Weyl semimetals, demonstrate that magnetostriction is a highly interesting probe for this class of materials and encourages further investigation.

This is the first time that we have achieved exceptionally good resolution using a PPMS system. Now, we reach the resolution limit of the best commercially available capacitance-measuring bridge (Andeen Hagerling 2550A), that is, the absolute value of the capacitance can be measured at a resolution of $10^{-6}$~pF. This demonstrates that the DynaCool system provides an extremely low-vibration environment for dilatometry measurements. However, to achieve such high-resolution measurements, it is necessary to set up the PPMS system in such a way that mechanical and, above all, electronic noise are avoided. The avoidance of ground loops is particularly important. This is discussed in detail in the next chapter.  

\section{\label{chap3a}Electronic isolation of the measurement setup}

Capacitance measurement is one of the most challenging types of electrical measurement. To perform high-resolution dilatomety measurements on our setup it is essential to provide a low-noise environment. In such measurements, the main sources of noise are stray capacitance between measurement wires and the surroundings, electrical noise due to the presence of ground loops, and noise from additional measurement electronics (measurement PC, magnet and temperature controllers, pumps, etc...)

\begin{figure*}
\subfloat{\label{figcsa}}
\subfloat{\label{figcsb}}
\subfloat{\label{figcsc}}
 \centering
 \includegraphics[width=\linewidth]{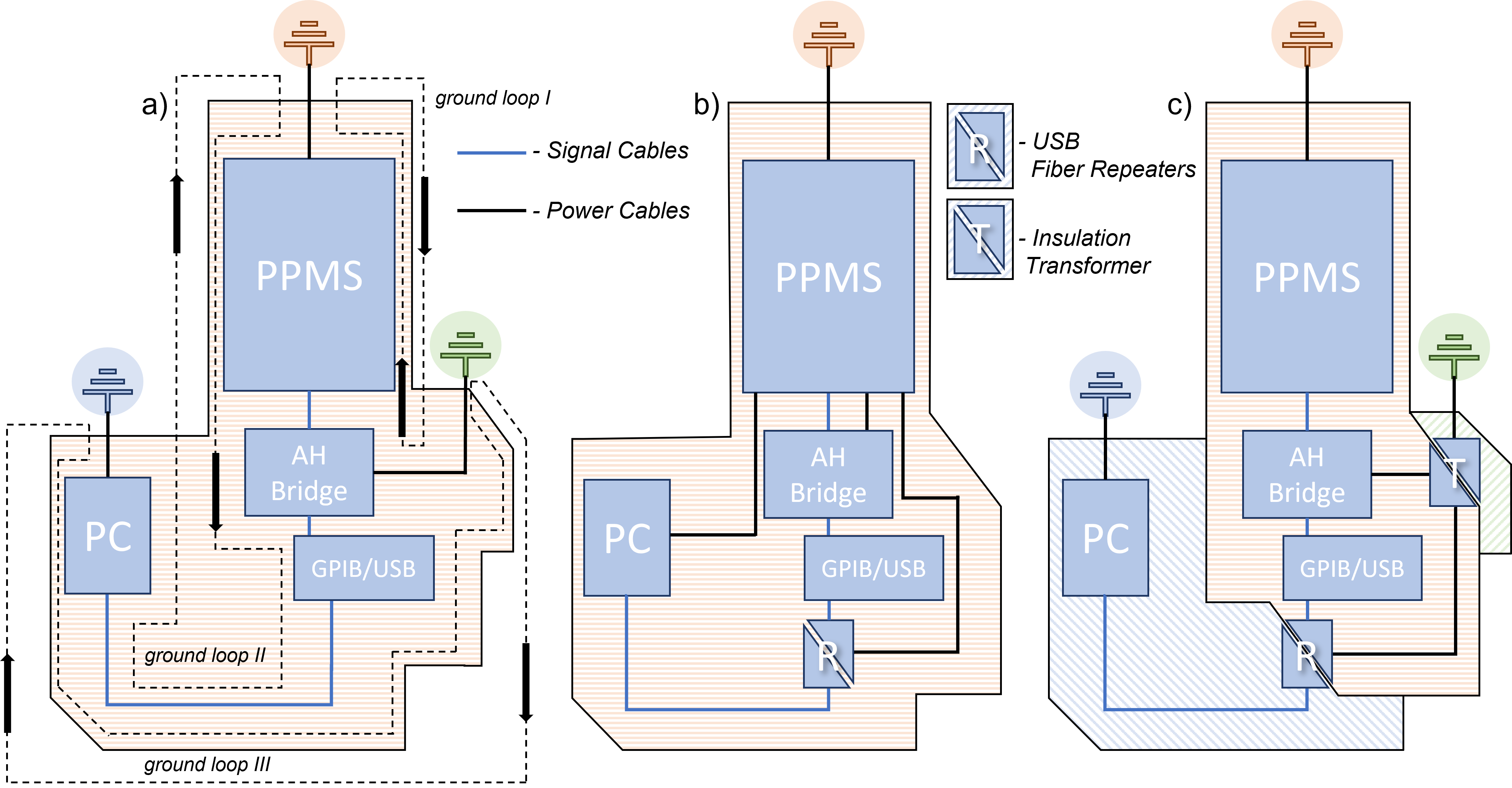}
 \caption{Three ways of electrical connection of the measuring system. a) An incorrectly connected circuit exposed to ground loops noise. b) The entire system is powered by a common PPMS power supply. c) Physical separation of the AH Bridge (Capacitance Bridge), PPMS (physical property measurement system including its controllers), and the PC (measuring computer).}
 \label{figcs}
\end{figure*}

In our setup, the noise originating from the stray capacitance is limited by the use of fully  shielded coaxial cables between the capacitor plates in the dilatometer and AH 2500 capacitance bridge. In addition, the outer shield of the coaxial cables was soldered to dilatometer screws and connected to the metal surface of the dilatometer cell. Because the contact springs of the dilatometry probe touch the inner chamber of the PPMS cooling channel, this arrangement ensures that the shielding of the measurement cable and PPMS sample chamber are in galvanic contact. This further decreases the influence of the stray capacitance because the PPMS itself acts as a part of the electric shielding. Although such a setup limits stray capacitance, it can potentially lead to the appearance of additional noise if the cryostat and capacitance bridge are connected to separate sockets,  creating a ground loop. Ground loops are a major source of noise. Similar problems occur if a galvanic connection exists between the measurement PC, cryostat, and capacitance bridge, \textit{ that is }, via USB or GPIB cables (see Fig. ~\ref{figcsa}).

To achieve a maximum resolution of ~$10^{-6}$~pF in our setup, it is necessary to address the problem of ground loops. Fig.~\ref{figcsb} displays a schematic arrangement of instruments that reduces the influence of ground loops, which can be used in most PPMS systems. All measurement instruments, including the measurement computer, were connected to the same PPMS power source. Although such a design does not fully remove the issue of ground loops, it limits their influence because all electronics are grounded to the same point. Although this design is very convenient to set up, it has the disadvantage that the measurement electronics are not galvanically decoupled from other electronic devices. This can result in a significant noise source if not properly designed (\textit{e.g.}, typical digital PC computers are often not optimized to provide a low-noise environment). In addition, such setup is not always feasible for cryostat systems other than PPMS.

The setup best suited for optimum resolution, and thus the highest measurement quality, which was used for our measurements on the QD Dynacool, is shown in Fig. ~\ref{figcsc}. Here, the capacitance bridge is grounded in the cryostat sample chamber and is electrically isolated from both the power grid and the measurement computer. In such a setting, all ground loops are avoided and grounding to the cryostat ensures that the sample chamber, wire shielding, and capacitance bridge have no potential differences. In this setup, galvanic isolation of the capacitance bridge from the power line is achieved using an isolating Transformer (ETTK 2500 - Isolating Transformer 230 VAC). A more challenging issue is the galvanic isolation of the measurement instrument and PC used for data recording. In our setup, this was achieved using an optical USB repeater. Here, AH2500 was connected via a USB/GPIB adapter to the Icron Ranger 2324 USB transmitter and transmitted via multimode optical fiber to the receiver and then via USB to the measurement PC. This mode of connection has been proven to provide a reliable connection that fully isolates the measurement rig from computer electronics. 

The measurement implementation is shown in Fig. ~\ref{figcsc}. Although it requires additional instrumentation, it provides a range of additional benefits: (1) it can be utilized on cryostat systems of any manufacturer, (2) the isolating transformer apart from disconnecting the ground acts as a low-pass filter rejecting any high-frequency components carried in the power lines, and (3) optically decoupling the measurement instrument and computer ensures that high-frequency digital noise and potential ill design of grounding in USB and GPIB do not influence the measurement results. 

\section{\label{chap5}Test measurements with the new mini-stress dilatometer}

To demonstrate the functionality of the new mini stress-dilatometer, we measured the thermal expansion of multiferroic TbMnO$_{3}$ under stress and almost zero pressure. The ferric character of TbMnO$_{3}$ results from the interplay between magnetic and electric degrees of freedom~\cite{tokura14}. The choice of the system was based on the fact that its response to even minute structural distortions should be strong as interactions in cases of orbital and magnetic degrees of freedom are significantly sensitive to interionic distances.

We prepared a cuboid sample ($l\times{}w\times{}d=1.83\times{}1.22\times{}0.72$ mm$^{3}$) cut from a rod grown using the floating-zone technique. Thermal expansion was measured along the longest achievable direction coinciding with the $(1,0,1)$ crystallographic direction.

\begin{figure}
 \centering
 \includegraphics[width=\linewidth]{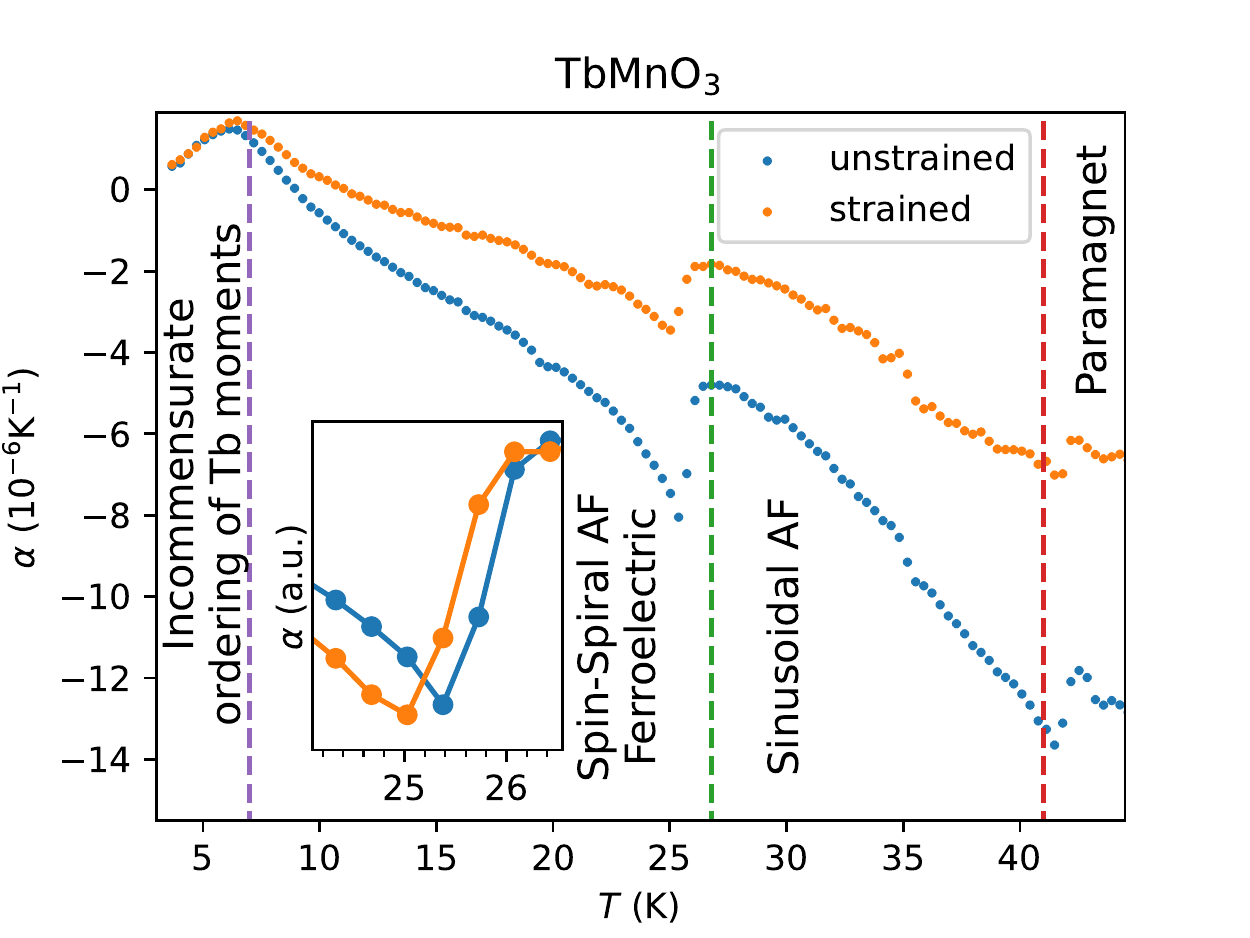}
 \caption{Thermal expansion coefficient of multiferroic TbMnO$_{3}$ measured using mini-dilatometers equipped with $2$ springs of $0.25$~mm (unstrained) and $4$ springs of $0.5$~mm. Dashed lines mark temperatures of the transitions observed in other works~\cite{tbmno3}. The inset shows the scaled and shifted region of data around the magnetic and para-ferroelectric transition. The solid line is a guide to the eye.}
 \label{figTbMnO3}
\end{figure}

At ambient pressure, TbMnO$_{3}$ undergoes a series of transitions between $42$ and $7$~K. Hallmarks of all those transitions can be found in the $T$-dependence of the thermal expansion coefficient, measured with a mini-dilatometer (see the blue curve in Fig. ~\ref{figTbMnO3}).  At $T_{\textrm{N\,{}Mn}}=42$~K, the Mn magnetic moments are arranged in an incommensurate sinusoidal antiferromagnetic structure with a temperature-dependent ordering wavevector $(0,k_{\textrm{Mn}}(T),0)$. Upon further cooling, a jump-like change in $\alpha$ occurs at $T\approx{}34$~K, which has been observed in previous studies and speculatively assigned to the alteration of the rate of change in the ordering wavevector as a function of temperature~\cite{tbmno3}. The most pronounced anomaly in the thermal expansion coefficient appears just below $26$~K and is a manifestation of the onset of an incommensurate cycloidal order of magnetic moments with the same non-zero components of the ordering wavevector as the sinusoidal structure. The cycloidal order breaks the inversion symmetry and allows for spontaneous electric polarization. This transition coincides with the onset of the ferroelectric phase with nonzero electric polarization $P\parallel{}c$~\cite{tbmno3}. Owing to its prominence, our investigation using uniaxial pressure will focus further on this feature. The last inflection in $\alpha$ observed above $5$~K results from the ordering of the magnetic moments residing at the Tb ions. 

The measurements under stress were performed using a new mini stress dilatometer. Here, the sample is clamped between the movable and fixed plate by four springs of 0.5 mm thickness, which exert a strong force of approximately 65 N. We measured a capacitance of 24 pF at 50 K, which changed slightly up to 25 pF during cooling to 2 K. By using the calibration curve of F(N) vs. C(pF) (see Fig. ~\ref{fig5}), we were able to determine a force of approximately 65 N applied to the sample. The applied stress was calculated by dividing the force by the cross section of the sample. Because we used a single rectangular crystal with the same cross section at both ends, a uniform stress along the sample was guaranteed, which is an important advantage compared to piezoelectric devices. The cut sample had a cross section of $w\times{}d=1.22\times{}0.72$ mm$^{2}$. This results in a uniaxial pressure of 75 MPa. In contrast, a force of only 4 N was applied to the sample during the mini-dilatometer measurements (see Fig. ~\ref{fig5}), which corresponds to 16 times lower stress. As can be seen in the orange curve in Fig. ~\ref{figTbMnO3}, an identical signature of four-phase transitions at 7 K, 26 K, 34 K and 42 K is also observed when measured under stress; only the absolute values are slightly smaller. Looking more closely at the most pronounced transition at 26 K, we note a shift in the transition temperature under stress towards lower temperatures (see the inset of Fig. ~\ref{figTbMnO3}). To accurately determine the transition temperature, we measured very slowly at a rate of 0.01 K/min. At this low sweeping rate, the measurements during heating and cooling overlapped very well. This shift in the transition temperature under stress toward lower temperatures is consistent with our calculations based on the Ehrenfest relationship. In the case of second-order phase transition, the Ehrenfest relation presented in Eq. ~\ref{ehrenfest} allows the calculation of the dependence of the phase transition temperature $\Delta{}T_{\textrm{C}}$ on the discontinuities in the volume thermal expansion and specific heat. A similar relationship holds for linear thermal expansion and allows the estimation of uniaxial pressure dependencies ~\cite{5,6,7}. Using the jump height at 26 K of our unstrained thermal expansion measurement and the jump height of the specific heat from Ref. ~\onlinecite{oflynn14}, we calculated the value of $\Delta{}T_{\textrm{C}}=-0.11$~K for the shift in the transition temperature, considering a uniaxial pressure increase of 70 MPa. The value measured by our strain dilatometry was $\Delta{}T_{\textrm{C}}=-0.3$~K (see the inset of Fig. ~\ref{figTbMnO3}) and agrees in magnitude and sign but is slightly larger. This observation is in agreement with similar studies of other systems ~\cite{35b}. Calculations using the Ehrenfest relation based on the determined step height of the linear thermal expansion coefficient appear to be suitable for estimating the uniaxial pressure dependence of $T_{\textrm{C}}$, but experimental measurements are required to determine the exact value. Errors can also occur in this context. Owing to the small shift in the temperature, the biggest challenge is to accurately determine the transition temperature. In a similar future study, we will attempt to measure thinner samples to increase the applied stress to achieve a greater shift in the transition temperature.

\section{\label{chap4}New application: Use inside a Dilution Refrigerator insert for the PPMS}

\begin{figure}
\subfloat{\label{fig8a}}
\subfloat{\label{fig8b}}
 \centering
 \includegraphics[width=\linewidth]{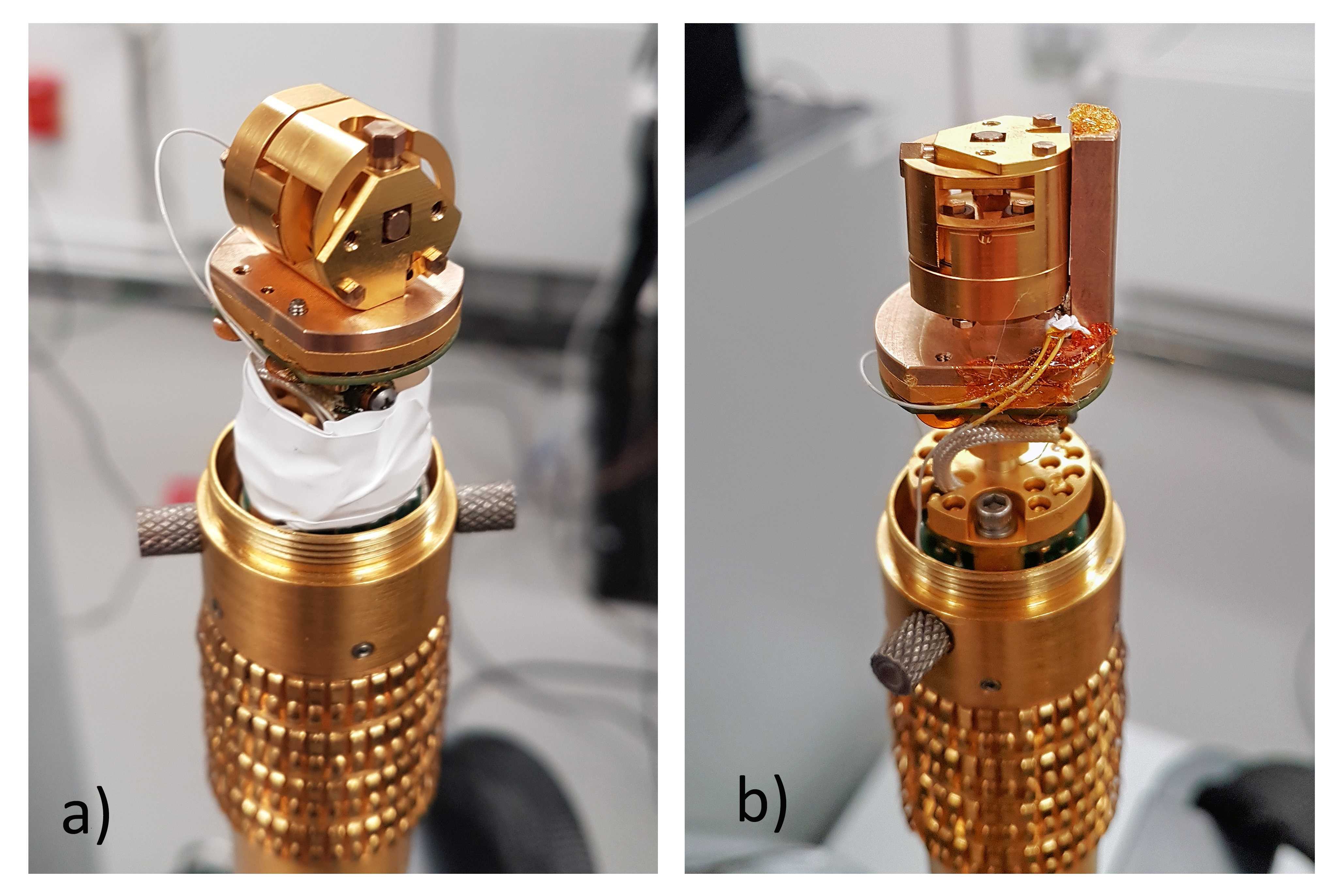}
 \caption{Photograph of the mini-dilatometer mounted on an adopter of a dilution refrigerator (a) perpendicular and (b) parallel to the applied magnetic field.}
 \label{fig8}
\end{figure}

The dilution refrigerator (DR) insert for the PPMS (DynaCool (D850) / PPMS (P850)) enables access to a temperature range spanning $4$~K down to $0.05$~K for a number of measurement options as well as for custom user experiments. Dilatometry measurements require the use of coaxial cables to limit stray capacitance. The standard Quantum Design dilution refrigerator (DR) unit was equipped with only a set of manganin DC lines. Thus, to perform dilatometric measurements, DR must be modified by adding coaxial cables. In the appendix, we explain in detail how we accomplished this. The mini- and mini-stress dilatometer can be screwed onto the DR insert experimental platform using two adapters made of Cu-Be. They were designed such that the change in the length of the sample could be measured parallel and perpendicular to the magnetic field (see Fig. ~\ref{fig8}). With this new setup, dilatometric measurements within the DR insert for the PPMS were possible for the first time. Using this setup, we cooled the dilatometers from room temperature to 65 mK within 8 h using this setup. After reaching this lowest temperature, the dilatometers were in thermodynamic equilibrium and ready for measurements within a few minutes.  

In the following section, we demonstrate the exceptional sensitivity of our mini-dilatometers when used within a DR insert in a PPMS DynaCool system. For this purpose, we show both low-temperature magnetostriction and thermal expansion measurements of a YbAlO$_{3}$ single crystal measured along the $c$-axis with a length of $l=1.74$~mm. Magnetostriction was measured using the configuration shown in Fig. ~\ref{fig8a}. Here, the change in length along $c$ was measured while the magnetic field was applied perpendicularly, \textit{i.e.}, along $a$.

\begin{figure}
\subfloat{\label{fig9a}}
\subfloat{\label{fig9b}}
 \centering
 \includegraphics[width=\linewidth]{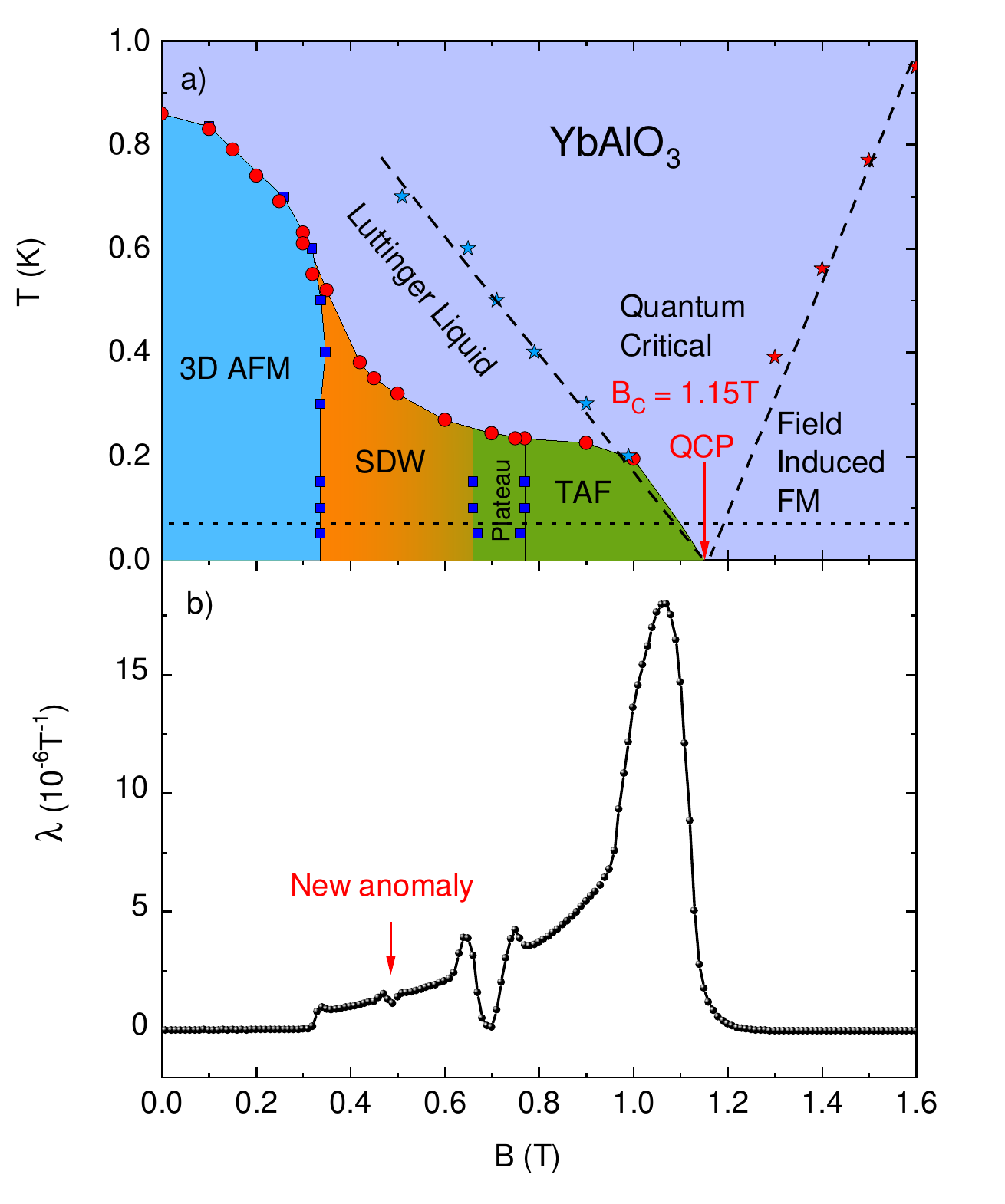}
 \caption{a) Phase diagram of YbAlO$_{3}$ reconstructed from specific-heat and magnetization measurements for $B\parallel{}a$, taken from Ref.~\onlinecite{37}. Blue, orange, and green colored areas show the three magnetically ordered phases. Blue squares at low temperatures and $B\approx{}0.7$~T within the green area marks the position of the $m=\frac{1}{3}$ plateau. b) Magnetostriction coefficient $\lambda (B)$, measured on a $1.74$~mm large single crystal along the $c$-axis at $T=0.065$~K. The magnetic field was applied along the $a$-axis.}
 \label{fig9}
\end{figure}

YbAlO$_{3}$ is described as a one-dimensional(1D) Heisenberg antiferromagnet spin $S=\frac{1}{2}$ chain, which shows Tomonaga–Luttinger liquid behavior at low temperatures and small magnetic fields~\cite{36}. Below the N\'{e}el temperature $T_{\textrm{N}}=0.88~$~K, finite dipolar interchain coupling with an Ising-like anisotropy causes a commensurate 3D-AFM order~\cite{37}. The B-T- phase diagram is shown in Fig. ~\ref{fig9}(a). The combination of isotropic intra- and Ising-like interchain interactions gives rise to the consecutive formation of a spin density wave (SDW) at $B=0.32$~T and a transverse antiferromagnetic~(TAF) phase as the magnetic field is increased~\cite{37}. Moreover, the magnetization curve exhibits a weak plateau close to $m=\frac{1}{3}$ in the field range of $B = 0.67$ to $B=0.76$~T. This feature is also included in the phase diagram shown in Fig. ~\ref{fig9}(a)~\cite{37}.

Fig.~\ref{fig9}(b) shows the magnetostriction coefficient $\lambda = 1/L (d\Delta L(B)/dB)$ measured at $T=0.065$~K. Owing to the excellent measurement resolution, even with a very small selected derivative interval of $dB = 0.01$~T, the curve of $\lambda(B)$ is almost noise-free and smooth. We can resolve all phase transitions identified thus far with the highest precision. Moreover, we detected clear double peaks at $B = 0.67$ and $B=0.76$~T. These are the boundary fields within which a weak plateau close to $m=\frac{1}{3}$  is observed in the magnetization curve. In addition, we found a small new anomaly previously found exclusively in thermal conductivity measurements ~\cite{38}. This clearly indicates the presence of another field-induced transition at approximately 0.5 T. This is discussed in more detail in ~\cite{38}.

\begin{figure}
\subfloat{\label{fig10a}}
\subfloat{\label{fig10b}}
 \centering
 \includegraphics[width=\linewidth]{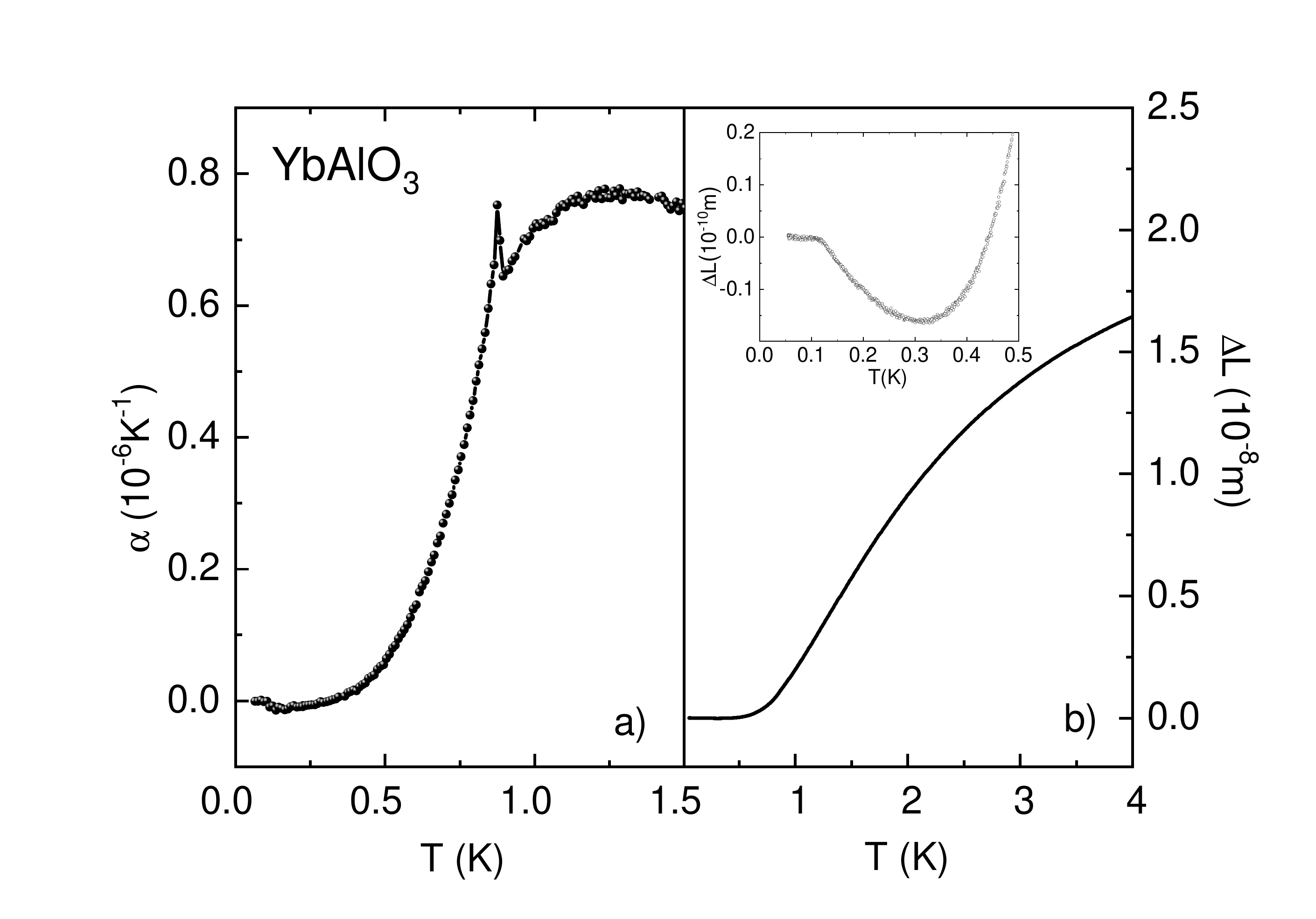}
 \caption{(a) The very sharp $\lambda$-type anomaly at $T_{\textrm{N}}=0.88$~K of the second order phase transition demonstrates the high sensitivity of our dilatometer and the very good quality of the sample. The inset of (b) demonstrates the very high resolution of our dilatometers of $\Delta L = 0.01$~\AA, when mounted on a DR-insert of a PPMS DynaCool system.}  
 \label{fig10}
\end{figure}

Fig.~\ref{fig10}(a) shows the low-temperature linear thermal expansion coefficient $\alpha=L^{-1} (dL/dT)_p$ and Fig.~\ref{fig10}(b), the absolute length change $\Delta{}L(T)$ of the YbAlO$_{3}$ single crystal measured along the $c$-axis in zero field. 

The sharp $\lambda$-type anomaly observed in $\alpha(T)$ at $T_{\textrm{N}}$ indicated a second-order phase transition. In addition, because the noise level was extremely low, we used a very narrow temperature window of $dT =10$ mK to determine the linear thermal expansion coefficient $\alpha=L^{-1} (dL/dT)_p$. The peak was very pronounced, and the transition occurred within a very narrow temperature interval of 50 mK. This shows that owing to the low mass of the dilatometer, thermal equilibrium was reached very quickly both within the dilatometer and within the sample. A complicated dilatometer setup often employed at low temperatures, in which the sample is thermally isolated from the cell, is no longer necessary with a tiny cell design. The inset shows the extraordinary sensitivity of our dilatometer with a very high resolution of $0.01$~\AA.

\begin{acknowledgments}
We are especially grateful to P. Gegenwart. Without his support and collaboration in the joint German Science Foundation Project No. KU 3287/1-1 and No. GE 1640/8-1 (tuning frustration in spin liquids by uniaxial pressure), the development of stress and mini-dilatometers would not have been possible. Gegenwart led this project, which also produced the uniaxial pressure results for CeRhSn. We also thank the founding director of our institute, Frank Steglich, who has always been a great advocate of dilatometry and supported us in obtaining our dilatometer patents. We would also like to thank M. Brando and A. P. Mackenzie for their support. Brando instigated the measurements of YbAlO$_{3}$. We thank L. Vasylechko and S. Nikitin for providing the YbAlO$_{3}$ single crystal. We are also indebted to T. Lühmann, who continuously improved the software required to run the experiments. We also thank the mechanics of our workshop, who made all the precise parts of the dilatometer and the PPMS-probe. Many thanks also to W. Schnelle and R. Koban, who had the idea and had already performed excellent angle-dependent magnetostriction measurements on NbP single crystals using our in-situ probe in an EverCool PPMS. We also thank V. Hasse and C. Shekhar for providing the NbP single crystals. In addition, we thank K. Povarov for advice on the installation of additional coaxial wires on the dilution unit. We would also like to thank T. Kubacka for sharing with us sample of TbMnO$_{3}$.
\end{acknowledgments}

\section*{Data Availability Statement}

The data supporting the findings of this study are available in this article.

\appendix
\section{Installation of the mini dilatometer in the Quantum Design dilution refrigerator}

Dilatometry measurements require the use of coaxial cables to limit stray capacitance. The standard Quantum Design dilution refrigerator (DR) unit was equipped with only a set of manganin DC lines. Thus, to perform dilatometric measurements, DR must be modified by adding coaxial cables. 

\begin{figure}
\includegraphics[width=0.6\linewidth]{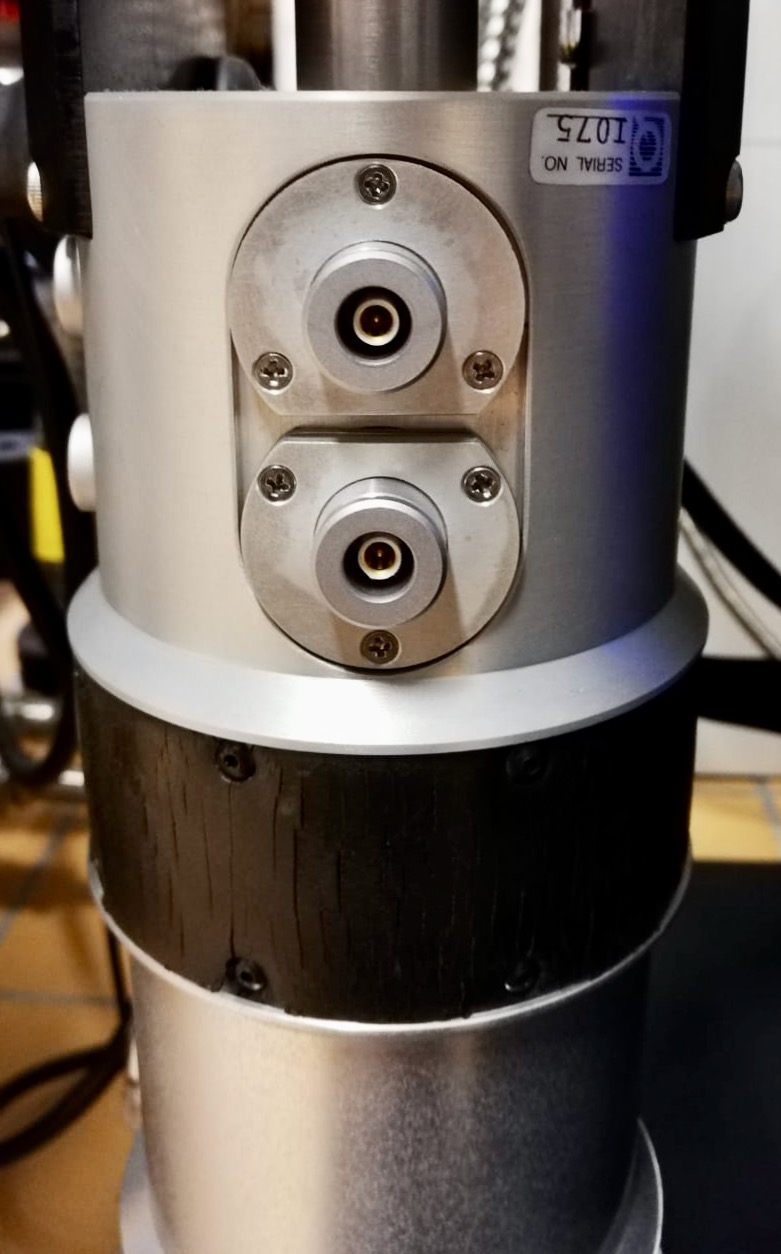}
 \caption{Custom made connectors.}
 \label{fig14b}
\end{figure}

To install the coaxial cables, we used spare signal conduits of the DR. They consisted of two Cu-Ni tubes that ran from the top of the unit down to the sample space. To be able to connect the coaxial cables from the outside, we replaced the aluminum blind flanges with new milled aluminum flanges with built-in vacuum-tight Fischer connectors (see Fig. ~\ref{fig14b}). To limit the heat load from the additional cables, we used a very thin 9450 WH Alpha Wire micro-coax cable with 150 $\mu$m cable outer diameter. In total, we used 1.5 meter long coaxial cables twice for wiring.

\begin{figure}
\includegraphics[width=0.95\linewidth]{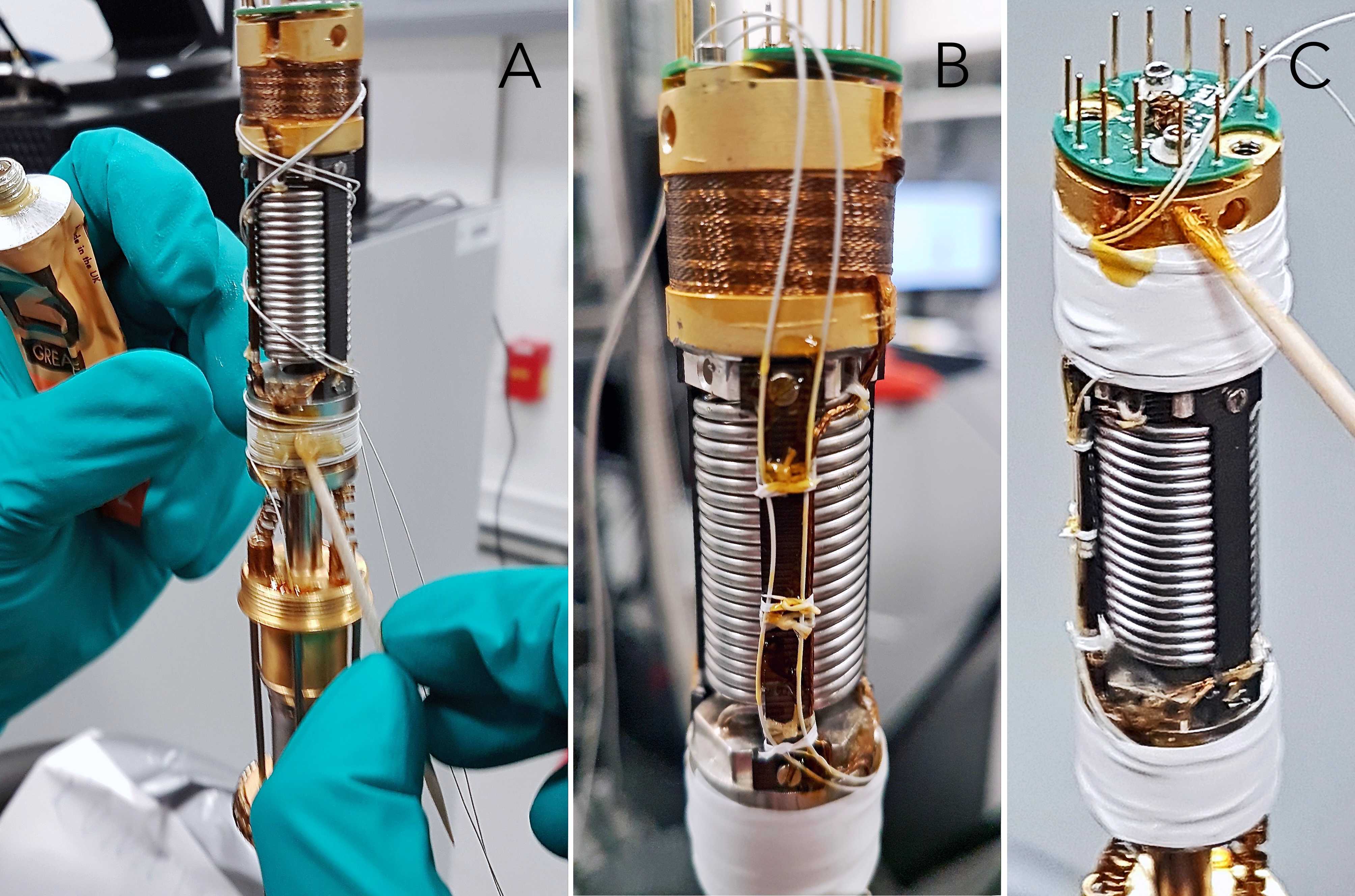}
 \caption{(a) Thermal anchoring of the coaxial wire to the still with GE-varnish. (b,c) Attachment of the coaxial wires to the vertical section of the DR using Teflon tape.}
 \label{fig15b}
\end{figure}

A crucial issue for the reliable operation of our setup at dilution temperatures was the proper thermalization of the signal cables before feeding them to the sample stage. This was achieved by the thermal anchoring of the wires to the probe stick by wrapping it around the central rod at the point where the wires exit the space cable conduits (see Fig. ~\ref{fig15b}(a,b)). To attach the wires, we used thermal varnish. Subsequently, the wires were fed vertically across the still, heat exchangers, and mixing chamber up to the sample stage. Here, an important point is that the wire must be firmly attached so that it does not touch the condenser tube (see Fig. ~\ref{fig15b}(b,c)). To accomplish this, we used GE-varnish and Teflon tape. 

\begin{figure}
\includegraphics[width=0.95\linewidth]{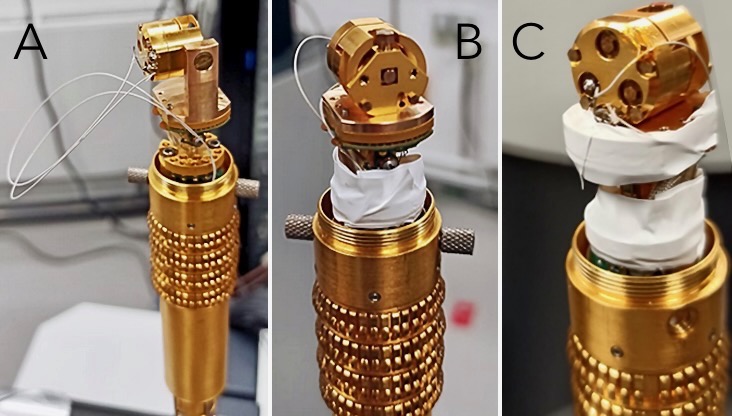}
 \caption{Different views of the final assembly. Teflon tape is used to handle excess wiring.}
 \label{SI2}
\end{figure} 

\begin{figure}
\includegraphics[width=0.7\linewidth]{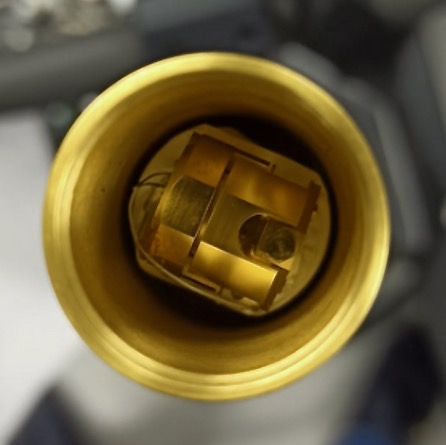}
 \caption{Top view of the final assembly with installed condenser tube.}
 \label{SI3}
\end{figure}

The final dilatometer assembly of the DR unit is illustrated in Fig. ~\ref{SI2}. In our setup, the wires were soldered directly to the dilatometer for each installation. The spare wire was wrapped around the sample stage using Teflon tape to prevent contact with the condenser tube. In the next step, we will connect coaxial cables with female and male connectors to make the installation more convenient. A top view of the final assembly with the condenser tube is shown in Fig. ~\ref{SI3}.

\nocite{*}
\bibliography{main}

\begin{thebibliography}{46}
\bibitem [{5()}]{5}%
  \BibitemOpen
  \href@noop {} {}\bibinfo {note} {The relation
  $dT_{C}/dp_{i}=T_{C}\Delta{}\alpha_{i}/\Delta{}C_{i}$ holds~\cite{6}, despite
  recent doubts~\cite{7}, because in the experiment always all (different)
  uniaxial pressures are kept constant and $\Delta{}C_{i}$ thus refers not only
  to the heat capacity jump at constant pi but in fact to $\Delta{}C$ measured
  at constant hydrostatic pressure).}\BibitemShut {Stop}%
\end{thebibliography}

\end{document}